\documentclass[a4paper,11pt]{article} \pdfoutput=1

\usepackage{jheppub}
\usepackage[T1]{fontenc}
\usepackage{amssymb,amsmath}
\usepackage{braket,bbm,bm}
\usepackage{mathtools}
\usepackage[dvipsnames,svgnames,table]{xcolor}
\usepackage[utf8]{inputenc}
\usepackage{soul, color}
\usepackage{lmodern}
\usepackage{footnote}
\usepackage[normalem]{ulem}
\usepackage{glossaries-extra}
\usepackage{slashed}
\usepackage{multirow}
\usepackage{here}
\usepackage{hyperref}
\usepackage{verbatim}
\usepackage{cleveref}
\usepackage{listings}
\usepackage{fancyvrb}
\usepackage{dsfont}
\usepackage{nicefrac,xfrac}
\usepackage{bbm}

\Crefname{equation}{eq.}{eqs.} \Crefname{section}{section}{sections} \Crefname{figure}{figure}{figures} \Crefname{appendix}{appendix}{appendices}

\let\vec\undefined \newcommand{\vec}[1]{\boldsymbol{#1}} \let\Re\undefined \let\Im\undefined \DeclareMathOperator{\Re}{Re} \DeclareMathOperator{\Im}{Im} \DeclarePairedDelimiter\abs{\lvert}{\rvert} \DeclareMathOperator{\arccosh}{arccosh}

\definecolor{jlab_red}{RGB}{192,39,45} \definecolor{jlab_orange}{RGB}{249,102,0} \definecolor{jlab_blue}{RGB}{47,122,121} \definecolor{jlab_green}{RGB}{65,125,10}

\hypersetup{
pdftitle = {Complex-momentum form factors from lattice QCD}, pdfsubject = {Complex-momentum form factors from lattice QCD}, pdfkeywords = {lattice QCD, form factors, imaginary momenta}, pdfauthor = {Maxwell T. Hansen, Gurtej Kanwar}, pdfnewwindow = {true}, colorlinks = {true}, linkcolor = {jlab_blue}, citecolor = {jlab_blue}, filecolor = {jlab_blue}, urlcolor = {jlab_blue}}

\title{\boldmath Complex-momentum form factors from lattice QCD}

\author[a,b]{Maxwell T.~Hansen} \author[a,b]{and Gurtej Kanwar}

\affiliation[a]{School of Physics and Astronomy, University of Edinburgh, Edinburgh EH9 3JZ, UK} \affiliation[b]{Higgs Centre for Theoretical Physics, School of Physics and Astronomy, The University of Edinburgh, James Clerk Maxwell Building, Peter Guthrie Tait Road, Edinburgh, EH9 3FD}

\emailAdd{maxwell.hansen@ed.ac.uk} \emailAdd{gkanwar@ed.ac.uk}

\abstract{We demonstrate that form factors evaluated with momenta analytically continued into the complex plane can be accessed directly from position-space Euclidean correlators such as those computed numerically in lattice quantum field theory, without continuing the correlators themselves or solving an inverse problem. The method applies a bilateral Laplace transform to lattice data, yielding the form factor throughout the transform's domain of convergence. For the pion form factor in Quantum Chromodynamics (QCD), this domain contains all complex values of the invariant $Q^2$ satisfying $\Re Q^2>-4m_\pi^2$, where $m_\pi$ denotes the pion mass. This extended kinematic reach could improve charge-radius determinations and provide more complete input for fits based on conformal mappings. After developing the method for infinite-volume correlators, we discuss several finite-volume estimators suitable for realistic calculations, including position-space fits and the use of a spatial chemical potential implemented through boundary conditions with an imaginary twist angle. Finally, we test the method in the two-dimensional $O(3)$ nonlinear sigma model using Monte Carlo data.}

\begin{document} \maketitle \flushbottom

\section{Introduction}

Lattice-QCD determinations of pion form factors, nucleon axial form factors, and nucleon electromagnetic form factors have reached a high level of precision over the past decade, enabling controlled studies of systematic effects and phenomenologically relevant observables such as axial couplings and charge radii~\cite{Brandt:2012zza,Brandt:2013dua,Gulpers:2013uca,Capitani:2015sba,Capitani:2017qpc,Djukanovic:2019jtp,Djukanovic:2021cgp,Djukanovic:2022wru,Djukanovic:2023beb,Barone:2025rye}. Precise determinations of nucleon electromagnetic form factors, for example, are also motivated by the long-standing proton-radius puzzle, where extractions from muonic hydrogen and from electronic probes initially led to significantly different values~\cite{Carlson:2015jba}. Although subsequent experimental and theoretical developments have clarified parts of this picture~\cite{Karr:2020wgh,Peset:2021iul,Gao:2021sml}, the episode continues to highlight the importance of independent, systematically controlled determinations of such hadronic properties from QCD.

\begin{figure}[t] \centering \includegraphics{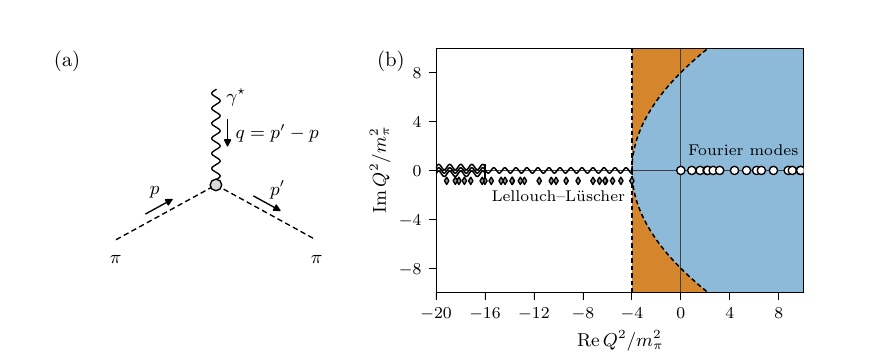} \caption{(a) The pion form-factor vertex, with incoming momentum $p$, outgoing momentum $p'$, and momentum transfer $q=p'-p$. (b) Domains accessible by complex-momentum Fourier transforms: general kinematics give the orange half-plane, while back-to-back kinematics give the blue parabolic subdomain. The timelike side is accessed using the Lellouch--L\"uscher relation; grey diamonds show the noninteracting two-pion finite-volume spectrum for $m_\pi L=6$ and total momenta $\boldsymbol P^2\leq 4(2\pi/L)^2$, transformed to the center-of-mass frame and offset vertically for visibility. White circles denote Fourier modes for various choices of incoming and outgoing pion spatial momenta, including pairs with different magnitudes.} \label{fig:summary} \end{figure}

A standard approach to lattice form-factor calculations in the spacelike region involves constructing three-point functions $\langle \mathcal{O}(\vec{p}') J^\mu(0) \overline{\mathcal{O}}(\vec{p}) \rangle$, where $\overline{\mathcal{O}}$ and $\mathcal{O}$ are the hadron source and sink operators, respectively, and $J^\mu$ is the current. The incoming and outgoing momenta are $\vec{p}$ and $\vec{p}'$, respectively, and the three-momentum transfer is $\vec{q}=\vec{p}'-\vec{p}$. The four-momentum transfer at the current determines the invariant $Q^2 \equiv \vec{q}^{\,2}-(q^0)^2$ of the extracted form factor, where $q^0$ is the difference between the outgoing and incoming on-shell hadron energies. Figure~\ref{fig:summary}(a) illustrates the setup for the pion vector form factor, the primary focus of this work. A central limitation in this approach is that finite-volume spatial momenta are discrete, so the form factor is typically sampled at a limited, discrete set of spacelike $Q^2 \geq 0$. This discrete set is illustrated in figure~\ref{fig:summary}(b) by the white circles at nonnegative $Q^2$, labeled ``Fourier modes.'' Quantities such as radii must then be inferred from fits to data near $Q^2=0$.

One way to overcome this limitation is to use twisted boundary conditions~\cite{Bedaque:2004kc,Sachrajda:2004mi,Tiburzi:2006px,Flynn:2007ess}. Instead of requiring the fields to be strictly periodic in space, one imposes a phase when a quark field is translated around the spatial volume. This shifts the allowed hadron momenta by an amount controlled by the twist angle, making it possible to tune the momentum transfer continuously between the usual Fourier modes. In many applications the twist can be applied to just the valence quarks without approximation, so new gauge ensembles are not needed for each momentum choice. Another alternative is the Rome method, in which correlation functions are differentiated with respect to external spatial momenta so that form factors and their derivatives can be accessed directly at $Q^2=0$~\cite{deDivitiis:2012vs,Hasan:2017wwt,Feng:2019geu}. In practice, the momentum derivatives are evaluated by differentiating quark propagators, leading to correlation functions with additional vector-current insertions from which the desired form-factor derivatives can be isolated.

For timelike kinematics above the particle-production threshold, the form factor can be accessed through Euclidean correlators of the form $\langle \mathcal{O}(\tau)J^\mu(0)\rangle$. In this case, the hadronic operator $\mathcal O$ is chosen to interpolate the relevant multi-hadron spectrum, for example, as products of individually momentum-projected pion operators. Extracting the form factor from these correlators is an inverse problem, but the finite volume makes it tractable. In a periodic spatial box of length $L$, the spectral decomposition $\langle \mathcal{O}(\tau)J^\mu(0)\rangle_L=\sum_n A_n(L)e^{-E_n(L)\tau}$ contains a discrete, gapped set of energies $E_n(L)$. A combined analysis using a varied operator basis can determine these energies and the associated matrix elements~\cite{HadronSpectrum:2009krc,Dudek:2012xn,Briceno:2016kkp,Ortega-Gama:2024rqx}. Known finite-volume relations can then convert them into infinite-volume amplitudes, including the timelike pion form factor, below the four-particle threshold~\cite{Lellouch:2000pv,Meyer:2011um,Briceno:2014uqa,Briceno:2015csa}. The accessible finite-volume levels are illustrated in figure~\ref{fig:summary}(b) by gray diamonds at negative $Q^2$.

Spectral function reconstruction gives an alternative approach for multi-hadron amplitudes, including timelike form factors. For a generic correlator $G(\tau)$ sampled at a discrete set of Euclidean times $\tau_i$, linear methods inspired by Backus--Gilbert reconstruction identify coefficients $c_i^\kappa(\bar\omega)$ such that $\sum_i c_i^\kappa(\bar\omega) e^{-\omega \tau_i}$ approximates a particular target smearing kernel $\kappa(\omega-\bar\omega)$. The corresponding linear combination $\sum_i c_i^\kappa(\bar\omega) G(\tau_i)$ then gives a controlled smeared estimate of the desired spectral or real-time quantity, rather than a pointwise observable~\cite{Backus:1968svk,Hansen:2019idp,Hansen:2017mnd,Bulava:2021fre}. Provided one can take the infinite-volume limit, followed by the zero-width limit of the smearing function $\kappa(\omega - \bar \omega)$, this method can be used to access a wide range of physical amplitudes, including the timelike pion form factor~\cite{Hansen:2019idp,Bulava:2021fre,Patella:2024cto,Evangelista:2023fmt,DiCarlo:2025mnm}. Formally, the method applies to all timelike kinematics.

In this work, we explore a new approach in which position-space Euclidean correlators are Fourier transformed with complex spatial momenta. The basic observation is that the exponential falloff of hadronic correlators guarantees convergence of a bilateral Laplace transform, i.e., a Fourier transform evaluated at complex momentum, within a finite domain. By isolating the analytically continued ground-state contribution through a plateau in an appropriate ratio of correlation functions at large Euclidean time, the usual form factor continued to complex momentum transfer can be extracted. In particular, imaginary spatial momenta provide access to real timelike $Q^2$ between the origin and the first particle-production threshold, while the general complex-momentum construction gives access to the half-plane bounded by the vertical line through the nearest particle-production branch point or stable-particle pole.\footnote{For a general form factor, the convergence region is fixed by the slowest decay among the possible asymptotic clusterings of the three operators; see section~\ref{subsec:accessible_domain}.}
Returning to figure~\ref{fig:summary}(b), the orange region indicates this full half-plane, while the blue parabolic region shows the subdomain accessible with back-to-back kinematics.

The bilateral Laplace transform provides a natural intermediary between standard Fourier-transform methods that hold for $Q^2 \geq 0$ and methods based on spectral-function reconstruction that hold for $Q^2 < - 4 m_\pi^2$. In all three cases, one forms the weighted combination $\sum_i w_i C_i$ of position-space Euclidean correlator data $C_i$, where $i$ is a generalized index that can label both spatial and Euclidean-time coordinates. However, the nature of the weights $w_i$ differs between the cases:
\begin{equation}
w_i =
\begin{cases}
e^{i \boldsymbol p\cdot \boldsymbol x_i} \, , & \text{standard Fourier transform, } Q^2 \geq 0 \, , \\[2mm] e^{i \boldsymbol k\cdot \boldsymbol x_i - \boldsymbol\kappa\cdot \boldsymbol x_i} \, , & \text{complex Fourier transform, } \Re Q^2 > -4m_\pi^2 \,, \\[2mm] c_i^\kappa(\bar\omega) \, , & \text{Backus--Gilbert reconstruction, } Q^2 < -4m_\pi^2 \, ,
\end{cases}
\end{equation}
Here $\boldsymbol k$ and $\boldsymbol\kappa$ are respectively the real and imaginary parts of the complex momentum; $\boldsymbol p=\boldsymbol k+i\boldsymbol\kappa$. Thus the spacelike case uses bounded Fourier phases, reconstruction methods use coefficients that are typically oscillatory in order to approximate the desired kernel, and the present approach uses complex Fourier weights, giving exponentially growing factors in some spatial directions that remain controlled by the exponential decay of the correlator within the convergence domain. We stress that, irrespective of the direction of $\boldsymbol\kappa$, the complex-momentum Fourier transform leads to a growing exponential factor in the half-space of $\mathbb R^3$ for which $\boldsymbol \kappa \cdot \boldsymbol x < 0$.

The discussion here and the more detailed treatment in section~\ref{sec:infinite_volume} below restrict attention to infinite-volume Euclidean correlators. Our perspective is to first establish that the idea is viable in the infinite-volume limit and then introduce finite-volume estimators that can be realistically computed using numerical lattice QCD. Formally, any quantity defines a viable estimator if it approaches the correct value as $L\to\infty$ and $T\to\infty$, where $T$ denotes the periodicity in the temporal direction. Going beyond this, it is instructive to identify improved choices that reduce systematic uncertainties at fixed $L$ and $T$. In section~\ref{sec:finite_volume} we discuss several possibilities. The simplest options involve direct truncated Fourier transforms or position-space fits and extrapolations. A more complicated option is to reconstruct part of the spectrum to extend the correlator in the spatial direction, as has been applied successfully to muon $g-2$~\cite{Gounaris:1968mw,Meyer:2011um,Bernecker:2011gh,Francis:2013qna,Aubin:2015rzx,DellaMorte:2017dyu,DellaMorte:2017khn,Giusti:2018mdh,Giusti:2019xct,Erben:2019nmx,Bruno:2019nzm,Bruno:2026yba,Borsanyi:2020mff,Boccaletti:2024guq,Lehner:2020crt,Ce:2022kxy,RBC:2023pvn,RBC:2024fic,Djukanovic:2024cmq,Bazavov:2024eou,Aliberti:2025beg,Lehner:2025qrl}. Finally, we discuss estimators based on applying a complex twisted boundary condition. To our knowledge this approach has not been well explored in lattice QCD, though related complex momenta and imaginary vector potentials have been considered in various condensed-matter and materials-science applications~\cite{Fukui:1998if,Hatano:1996ntg,Reuter2017ComplexBand,Jensen2017Eigenchannels,Zhong:2025nhb}.

Before closing this introduction, we note that a number of the ideas presented here overlap existing lattice-QCD-specific literature. Related position-space and coordinate-space strategies include:
\begin{itemize}
\item the time-momentum representation and long-distance spectral reconstruction methods used for the hadronic vacuum-polarization contribution to muon \(g-2\)~\cite{Bernecker:2011gh};
\item Euclidean-time-integral methods for the neutral-pion transition form factor \(\pi^0\to\gamma^\star\gamma^\star\)~\cite{Feng:2012pion2gamma,Meyer:2013pion2gamma,Lin:2025pionPole};
\item position-space formulations of hadronic light-by-light scattering for muon \(g-2\), where continuum QED kernels are combined with lattice-QCD four-point functions~\cite{Asmussen:2016lse};
\item thermal-photon-production calculations in which Euclidean correlators at imaginary spatial momentum are used to analytically continue the retarded current correlator to vanishing photon virtuality~\cite{Meyer:2018xpt,Torok:2022vki,Krasniqi:2024inr};
\item methods based on QED\(_\infty\) that combine finite-volume QCD data with infinite-volume photon kernels to remove power-law finite-volume effects~\cite{Feng:2018qpx};
\item master-field and stochastic-locality approaches that exploit localized position-space correlators in large-volume lattice QCD~\cite{Ce:2021akh,Bruno:2023vhs}.
\end{itemize}

What distinguishes the present work is both that the hadron itself, here the pion, carries complex-valued spatial momentum and, closely related to this, that the analytic continuation induced by complex spatial momenta is examined directly in the time--momentum representation of the correlator. In particular, our method isolates ground-state time dependence of the form $e^{- E_\pi(\boldsymbol p) \tau}$ in terms of the analytically continued energies $E_\pi(\boldsymbol p) = \sqrt{m_\pi^2 + (\boldsymbol k + i\boldsymbol{\kappa})^2}$. While this work focuses on applying this method to extract form factors, the concept of using complex momenta in the time--momentum representation is general and may prove useful in other contexts.

\bigskip

The remainder of this manuscript develops the infinite-volume construction in section~\ref{sec:infinite_volume}, examines finite-volume estimators in section~\ref{sec:finite_volume}, and tests the approach in the two-dimensional $O(3)$ nonlinear sigma model in section~\ref{sec:numerical_check}. We conclude in section~\ref{sec:conclusions}; two appendices provide additional details on position-space momentum derivatives and the accessible complex momentum-transfer domain.

\section{Analytically continued form factors from position-space correlators} \label{sec:infinite_volume}

We now formulate the complex-momentum construction in the infinite-volume continuum theory. Using the pion electromagnetic form factor as a concrete example, we first review the standard spacelike ground-state saturation relations and then determine the domain in which they continue to complex spatial momenta. The extension to other hadronic form factors is immediate once the relevant convergence boundary is identified.

\subsection{Spacelike momentum transfer}

We first briefly review the standard approach for accessing spacelike form factors from Euclidean correlators. This serves to establish notation and set a standard for comparison to our proposal.

The Euclidean two- and three-point functions of relevance to the pion form factor are
\begin{align}
\label{eq:Gx} \delta^{ab} G^{AB}(\tau,\boldsymbol{x}) & = \langle 0 \vert O^a_A(\tau,\boldsymbol{x}) O^b_B(0) \vert 0 \rangle\,,\\ \label{eq:Cxy} \epsilon^{abc} C^{AB}_{\mu}(\tau,t;\boldsymbol{x},\boldsymbol{y}) & = \langle 0 \vert O^a_A(\tau,\boldsymbol{x}) J_\mu^b(t,\boldsymbol{0}) O^c_B(0,\boldsymbol{y}) \vert 0 \rangle\,,
\end{align}
where we assume an infinite-volume continuum theory, and where $O^a_A(x)$ and $O^b_B(x)$ are the respective sink $\mathcal{O}$ and source $\overline{\mathcal{O}}$ interpolating operators with isospin indices $a,b$ and additional indices $A$, $B$ indicating the particular choice of operator, e.g.\ local or smeared. The operator $J_\mu^b(z)$ is the isovector vector current with isospin index $b$,
\begin{equation}
J_\mu^b(z) = \overline \psi(z) \gamma_\mu \frac{\tau^b}{2} \psi(z)\,,
\end{equation}
where $\psi = (\mathcal U, \mathcal D)^T$ is the isospin doublet of up and down quarks, $\tau^b$ is a Pauli matrix acting in isospin space, and $\gamma_\mu$ is a Euclidean gamma matrix. Here $\tau > t > 0$ so that the operators are time ordered, and we have explicitly factored out the isospin structure in terms of the Kronecker delta and antisymmetric Levi-Civita tensor. In both the two- and three-point functions we make use of translational invariance by setting one of the spatial coordinates and one of the time coordinates to zero, without loss of generality.

The standard method proceeds by transforming to the time--momentum representation by applying spatial Fourier transforms to the correlators, giving
\begin{align}
G^{AB}(\tau,\boldsymbol{p}) & = \int \!{\rm d}^3 \boldsymbol{x} \, e^{-i \boldsymbol{p} \cdot \boldsymbol{x}} G^{AB}(\tau,\boldsymbol{x})\,, \label{eq:G_FT_time_momentum} \\ C^{AB}_{\mu}(\tau,t;\boldsymbol{p}',\boldsymbol{p}) & = \int \!{\rm d}^3 \boldsymbol{x} \, e^{-i \boldsymbol{p}' \cdot \boldsymbol{x}} \int \!{\rm d}^3 \boldsymbol{y} \, e^{i \boldsymbol{p} \cdot \boldsymbol{y}} C^{AB}_{\mu}(\tau,t;\boldsymbol{x},\boldsymbol{y})\,. \label{eq:C_FT_time_momentum} \end{align}
For simplicity, we reuse the labels $G^{AB}$ and $C^{AB}_\mu$ in the time--momentum representation, indicating the momentum-space dependence through the arguments $\vec{p}$ and $\vec{p}'$.

To extract physical observables here, it is standard to next insert a complete set of states, in order to make the Euclidean-time dependence explicit. For fixed spatial momenta $\boldsymbol{p}$ and $\boldsymbol{p}'$, the lowest energy states accessed by the interpolating operators under consideration are single-pion states with fixed momentum $\vec{p}$ or $\vec{p}'$ and corresponding energy $E_\pi(\vec{p}) = \sqrt{m_\pi^2 + \vec{p}^2}$ or $E_\pi(\vec{p}') = \sqrt{m_\pi^2 + \vec{p}'^2}$. The relevant ground-state-saturated contributions to the two- and three-point functions are then given by
\begin{align}
G^{(0),AB}(\tau,\boldsymbol{p}) & = \frac{Z_\pi^A Z_\pi^B}{2 E_\pi(\boldsymbol{p})} e^{-E_\pi(\boldsymbol{p}) \tau}\,, \label{eq:G_tmr_saturation} \\ C_\mu^{(0),AB}(\tau,t; \boldsymbol{p}',\boldsymbol{p}) & = \frac{-i Z_\pi^A Z_\pi^B} {4 E_\pi(\boldsymbol{p}') E_\pi(\boldsymbol{p})} e^{-E_\pi(\boldsymbol{p}')(\tau-t)} e^{-E_\pi(\boldsymbol{p}) t} (p' + p)_\mu^{E} F_\pi(Q^2)\,. \label{eq:C_tmr_saturation}
\end{align}
These contributions dominate for large time separations $\tau \gg 1/m_\pi$ in the two-point function and $t \gg 1/m_\pi$, $(\tau-t)\gg 1/m_\pi$ in the three-point function. Here, the relativistic normalization of the single-particle states is used and the overlap factors are defined by $\langle 0 | O_A^a(0, \boldsymbol{x}) | \pi^b(\boldsymbol{p})\rangle = \delta^{ab} Z^A_\pi e^{i \boldsymbol{x} \cdot \boldsymbol{p}}$, where the continuum, infinite-volume overlap factors are evaluated on shell, $Z_\pi^A(p^2) = Z_\pi^A(-m_\pi^2) \equiv Z_\pi^A$ and $Z_\pi^B(p^2) = Z_\pi^B(-m_\pi^2) \equiv Z_\pi^B$, and are therefore independent of the kinematics.

In eq.~\eqref{eq:C_tmr_saturation} we have introduced the form factor $F_\pi(Q^2)$ defined via the matrix element
\begin{equation}
\langle \pi^a(\vec{p}') | J_\mu^b | \pi^c(\vec{p}) \rangle = -i \, \epsilon^{abc} (p' + p)_\mu^E F_\pi(Q^2)\,,
\end{equation}
in terms of the Euclidean four-momentum transfer \((p' + p)_\mu^{E} = (i E_\pi(\vec{p}') + i E_\pi(\vec{p}), \vec{p}' + \vec{p})\) and invariant
\begin{equation}
Q^2 = (\boldsymbol{p}' - \boldsymbol{p})^2 - (E_\pi(\boldsymbol{p}') - E_\pi(\boldsymbol{p}))^2\,. \label{eq:q2_definition}
\end{equation}
For any real spatial momenta, $Q^2$ is nonnegative, i.e.\ the form factor is evaluated in the spacelike region.

The ground-state-saturated expressions in eqs.~\eqref{eq:G_tmr_saturation} and \eqref{eq:C_tmr_saturation} are the leading terms in the full spectral representations of the correlators defined in eqs.~\eqref{eq:G_FT_time_momentum} and \eqref{eq:C_FT_time_momentum}. Decomposing the two- and three-point correlators fully into their ground-state terms and excited-state contamination gives
\begin{align}
G^{AB}(\tau, \boldsymbol{p}) &= G^{(0),AB}(\tau,\boldsymbol{p}) + \delta G^{AB}(\tau,\boldsymbol{p})\,, \label{eq:G_tmr_spectral} \\ C^{AB}_\mu(\tau,t;\boldsymbol{p}',\boldsymbol{p}) &= C_\mu^{(0),AB}(\tau,t;\boldsymbol{p}',\boldsymbol{p}) + \delta C^{AB}_\mu(\tau,t;\boldsymbol{p}',\boldsymbol{p})\,. \label{eq:C_tmr_spectral}
\end{align}
The excited-state contamination can be written as Laplace transforms of spectral densities with the pion pole removed,
\begin{align}
\delta G^{AB}(\tau,\boldsymbol{p}) &= \int_{m_\pi}^{\infty} {\rm d}M \, \rho^{AB}_{\slashed{\pi}}(M)\, \frac{e^{-E_M(\boldsymbol{p})\tau}}{2E_M(\boldsymbol{p})}\,, \label{eq:delta_G_spectral_density} \\ \delta C^{AB}_\mu(\tau,t;\boldsymbol{p}',\boldsymbol{p}) &= \int_{m_\pi}^{\infty} {\rm d}M' \int_{m_\pi}^{\infty} {\rm d}M \, \rho^{AB}_{\slashed{\pi},\mu}(M',M)\, \frac{ e^{-E_{M'}(\boldsymbol{p}')(\tau-t)} e^{-E_M(\boldsymbol{p})t} }{ 4E_{M'}(\boldsymbol{p}')E_M(\boldsymbol{p}) }\,, \label{eq:delta_C_spectral_density}
\end{align}
where \(E_M(\boldsymbol{p})=\sqrt{M^2+\boldsymbol{p}^2}\). The spectral densities \(\rho^{AB}_{\slashed{\pi}}(M)\) and \(\rho^{AB}_{\slashed{\pi},\mu}(M',M)\) are understood as distributions with the single- and double-pion poles subtracted, respectively, as indicated by the \(\slashed{\pi}\) subscript.

\subsection{Accessible domain of complex momentum transfer} \label{subsec:accessible_domain}

\begin{figure}[t] \centering \includegraphics{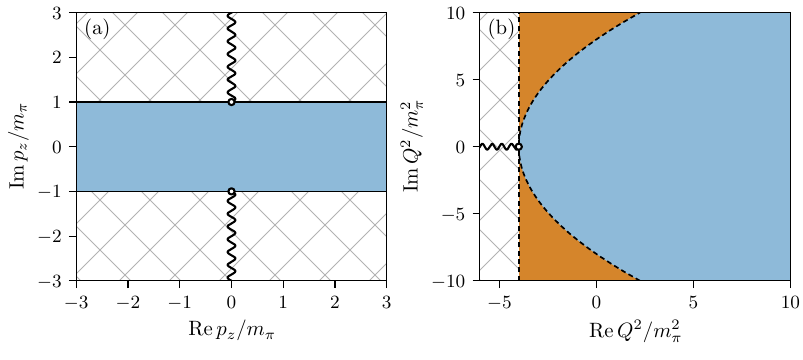} \caption{\textbf{(a)} Strip of convergence in the complex $p_z$ plane, after aligning the $z$ axis with the direction of the imaginary momentum. The solid black lines indicate $\Im p_z/m_\pi=\pm 1$, while the white circles and oscillating lines indicate the branch points and branch cuts extending away from the strip. \textbf{(b)} Accessible domain in the complex $Q^2$ plane. The orange background marks the full half-plane $\Re Q^2>-4m_\pi^2$ accessible with general incoming and outgoing momenta, while the light-blue overlay marks the parabolic back-to-back subdomain. The white circle and oscillating line indicate the branch point and branch cut along the real axis for $Q^2<-4m_\pi^2$. Faint cross-hatching marks the inaccessible regions in both panels.} \label{fig:Q2_domain} \end{figure}

We now determine the region of complex momenta in which the ground-state saturation relations in eqs.~\eqref{eq:G_tmr_saturation} and \eqref{eq:C_tmr_saturation} remain valid. Two conditions are required: the position-space transforms must converge without an inverse reconstruction, and their large-Euclidean-time behavior must be governed by states satisfying $p^2=p'^2=-m_\pi^2$. For back-to-back momenta, the resulting $Q^2$ values fill the parabolic region shown in figure~\ref{fig:Q2_domain}(b); general incoming and outgoing momenta enlarge this domain to the half-plane $\Re Q^2>-4m_\pi^2$, as demonstrated in appendix~\ref{app:q2_domain}.

Our result is based on a standard theorem concerning functions with exponential falloff~\cite{Paley:1934}. We state the theorem for a one-dimensional function \(f:\mathbb{R}\to\mathbb{R}\). This is sufficient for our construction because one can always align the coordinate system with the direction of the imaginary momentum, so that the essential setup is one dimensional. Suppose that \(f(x)\) falls off sufficiently quickly that, for some positive \(\lambda\),\footnote{More formally, \(f\) is exponentially weighted \(L^1\) with rate \(\lambda\), i.e., \(f \in L^1(e^{\lambda \abs{x}} dx)\).}
\begin{equation}
\int_{-\infty}^\infty dx \, e^{\lambda \abs{x}} \abs{f(x)} < \infty .
\end{equation}
The theorem then states that the Fourier transform
\begin{equation}
\label{eq:fourier-f} f(p) = \int_{-\infty}^\infty dx \, e^{i p x} f(x)
\end{equation}
converges and defines a holomorphic function for complex momenta \(p\) in the strip
\begin{equation}
\abs{\Im{p}} < \lambda .
\end{equation}
Equivalently, this holomorphic extension of the ordinary Fourier transform is the bilateral Laplace transform, with Laplace variable $s=-ip$.

Analyticity of the function inside the strip can be shown almost immediately. First, whenever \(\abs{\Im p}<\lambda\), the exponential factor in the Fourier transform cannot overcome the falloff of \(f\),
\begin{equation}
\abs{e^{i p x} f(x)} \leq e^{\abs{\Im{p}} \abs{x}} \abs{f(x)} ,
\end{equation}
ensuring that the integral in eq.~\eqref{eq:fourier-f} is absolutely convergent. Likewise any derivatives of this expression with respect to $\Re{p}$ and $\Im{p}$ generate harmless polynomial factors in \(\abs{x}\) that do not modify this region of absolute convergence. Derivatives with respect to \(p\) and its complex conjugate \(\bar p\) may therefore be interchanged with the integral. The Cauchy--Riemann condition then follows from
\begin{equation}
\frac{\partial}{\partial \bar{p}} f(p) = \int_{-\infty}^{\infty} dx \, \left( \frac{\partial}{\partial \bar{p}} e^{i p x} \right) f(x) = 0,
\end{equation}
since the integrand \(e^{i p x}\) is holomorphic in \(p\) for each fixed \(x\).

We now return to the physical system of interest, in particular to the Fourier transforms that lead to the time--momentum representation in eqs.~\eqref{eq:G_FT_time_momentum} and \eqref{eq:C_FT_time_momentum}. These Fourier transforms will converge for a strip of momenta $\boldsymbol{p}, \boldsymbol{p}' \in \mathbb{C}^3$ due to the exponential falloff of the position-space correlation functions $G^{AB}(\tau, \boldsymbol{x})$ and $C^{AB}_\mu(\tau, t; \boldsymbol{x}, \boldsymbol{y})$.

For the two-point function, the pion mass $m_\pi$ determines the exponential decay of the correlation function, leading to the region of convergence
\begin{equation}
\lVert \Im{\vec{p}} \rVert < m_\pi.
\end{equation}
The region of convergence for this case is shown schematically in figure~\ref{fig:Q2_domain}(a), with the labels indicating that the imaginary part of the momentum is aligned with the \(z\) axis.

For the three-point function, the exponent in the Fourier kernel depends on\linebreak $\vec{p} \cdot \vec{x} + \vec{p}' \cdot \vec{y}$. We are therefore interested in the directions with slowest exponential decay in the joint space $(\vec{x}, \vec{y})$. For the pion form factor under consideration here, the exponential falloff for asymptotically large $\lVert\vec{x}\rVert$ and $\lVert\vec{y}\rVert$ is $e^{-m_\pi(\lVert\vec{x}\rVert + \lVert\vec{y}\rVert)}$, regardless of direction, which immediately leads to convergence of the time--momentum correlation functions for all complex momenta $\vec{p}$ and $\vec{p}'$ satisfying
\begin{equation}
\label{eq:bilaplace-roc} \lVert \Im{\vec{p}} \rVert < m_\pi \quad \text{and} \quad \lVert \Im{\vec{p}'} \rVert < m_\pi.
\end{equation}
For other choices of external hadron or current, the convergence region must be determined by comparing the possible asymptotic clusterings of the three operators. When the two hadron interpolators remain together and are separated from the current, the falloff is governed by the lightest state with the quantum numbers of the current. The vector-current channel begins at $2m_\pi$ for isovector quantum numbers and at $3m_\pi$ for isoscalar quantum numbers. The isovector axial current contains a one-pion pole followed by a three-pion continuum, whereas the isoscalar axial current couples longitudinally to the $\eta$ and $\eta'$ and transversely to $f_1$ states, with its pure-pion continuum beginning at four pions. When either hadron interpolator is instead separated from the other two operators, the falloff is governed by the external hadron. Thus, for external pion states, the $m_\pi$ falloff associated with the pion operator determines the convergence region unless the current couples to a single pion or the vacuum.

While this demonstrates that eqs.~\eqref{eq:G_FT_time_momentum} and \eqref{eq:C_FT_time_momentum} are convergent integrals yielding complex-analytic functions within this strip of complex momenta, it remains to show two further points:
\begin{itemize}
\item The spectral representations in eqs.~\eqref{eq:G_tmr_spectral} and \eqref{eq:C_tmr_spectral} are complex analytic, and thus still hold, within the same region.
\item The analytic continuations of the ground-state saturation expressions in \eqref{eq:G_tmr_saturation} and \eqref{eq:C_tmr_saturation} still asymptotically dominate for $\tau \gg t \gg 1/m_\pi$.
\end{itemize}
We start by analytically continuing the values of the energies to complex momenta. Using relativistic invariance, the continuum, infinite-volume energies in the spectral representations above still satisfy the dispersion relations $E_M(\vec{p}) = \sqrt{M^2 + \vec{p}^2}$ and $E_{M'}(\vec{p}') = \sqrt{M'^2 + \vec{p}'^{\,2}}$, where $M$ and $M'$ are center-of-mass energies appearing in the spectral densities and $\vec{p}$ and $\vec{p}'$ are now complexified momenta. Importantly, we take the principal root in these expressions, so that the branch cut is placed along the negative real axis of the argument. With this choice, $E_M(\vec{p})$ and $E_{M'}(\vec{p}')$ are holomorphic within the region of convergence set by eq.~\eqref{eq:bilaplace-roc} and we have successfully analytically continued the spectral representations in eqs.~\eqref{eq:G_tmr_spectral} and \eqref{eq:C_tmr_spectral} into this domain. Next, using the principal root, the real part of the continued energies takes the form
\begin{equation}
\label{eq:real_part_continued_energy}
\begin{aligned}
\Re E_M(\vec{p}) &= \sqrt{ \frac{ \sqrt{\mathcal A_M(\vec{p})^2+\mathcal B(\vec{p})^2} +\mathcal A_M(\vec{p}) }{2} }\,, \\ \mathcal A_M(\vec{p}) &\equiv M^2+\lVert \Re \vec{p} \rVert^2-\lVert \Im \vec{p} \rVert^2\,, \\ \mathcal B(\vec{p}) &\equiv 2\,\Re \vec{p}\cdot\Im \vec{p}\,.
\end{aligned}
\end{equation}
Since $\Re E_M(\vec{p})$ increases with the center-of-mass energy $M$ on this sheet, the real parts of the analytically continued energies maintain their ordering. The same statement applies to $E_{M'}(\vec{p}')$ in the second spectral variable. As a result, for $\tau \gg t \gg 1/m_\pi$ the analytic continuations of the ground-state-saturated expressions \eqref{eq:G_tmr_saturation} and \eqref{eq:C_tmr_saturation} give the correct ground-state saturation of the analytically continued spectral representation. More explicitly, the decompositions and residual spectral representations in eqs.~\eqref{eq:G_tmr_spectral}--\eqref{eq:delta_C_spectral_density}, together with the monotonicity displayed in eq.~\eqref{eq:real_part_continued_energy}, show that every excited-state contribution continues to decay faster in Euclidean time than the corresponding ground-state term. Thus the hierarchy between the ground and excited states is preserved under the analytic continuation.

Focusing in particular on the form factor in \eqref{eq:C_tmr_saturation}, these arguments demonstrate that within the strip of convergence, the complex-valued Fourier transform of the three-point function can be related to the form factor $F_\pi(Q^2)$ evaluated at complex spatial momenta by extracting and removing the appropriately continued exponential and overlap factors. The squared momentum transfer is determined by evaluating $Q^2 = (\vec{p}' - \vec{p})^2 - \left( E_\pi(\vec{p}') - E_\pi(\vec{p}) \right)^2$ at the associated complex momenta.

For the particularly useful choice of back-to-back momenta $\boldsymbol p=-\boldsymbol p'=\boldsymbol k+i\boldsymbol\kappa$, where $\boldsymbol k,\boldsymbol\kappa\in\mathbb{R}^3$, one finds
\begin{equation}
Q^2 = 4\vec{p}^2 = 4(\boldsymbol k^2-\boldsymbol\kappa^2) +8i\,\boldsymbol k\cdot\boldsymbol\kappa\,.
\end{equation}
Writing $X=\Re Q^2$ and $Y=\Im Q^2$, the Cauchy--Schwarz inequality gives $Y^2\leq64\boldsymbol k^2\boldsymbol\kappa^2=16\boldsymbol\kappa^2(X+4\boldsymbol\kappa^2)$. The right-hand side is maximal as $\boldsymbol\kappa^2\to m_\pi^2$, so the exact back-to-back domain is
\begin{equation}
(\Im Q^2)^2<16m_\pi^2(\Re Q^2+4m_\pi^2)\,.
\end{equation}
This is the interior of a parabola in the complex $Q^2$ plane opening to the right, with vertex at $Q^2=-4m_\pi^2$, as illustrated in figure~\ref{fig:Q2_domain}(b). Appendix~\ref{app:q2_domain} gives the detailed derivation and an explicit construction reaching every point in this domain; it also shows that general $\boldsymbol p$ and $\boldsymbol p'$ enlarge the accessible region to the full half-plane $\Re Q^2>-4m_\pi^2$.

The utility of accessing this entire complex-momentum domain remains to be explored. One possibility is that it can further improve constraints on the charge radius by providing access to additional points in the complex $Q^2$ plane near $Q^2 = 0$. A second is that it may be useful in the context of parametrizations of the form factor, such as the $z$-expansion, which exploit analyticity properties in the complex $Q^2$ plane. A third possibility is that it may allow for the numerical construction of dispersive representations of the form factor to access regions near the branch cut.

\subsection{\texorpdfstring{Isolating \(F_\pi(Q^2)\)}{Isolating Fpi(Q2)}}

The most direct construction uses back-to-back momenta, $\boldsymbol p'=-\boldsymbol p$. Then $E_\pi(\boldsymbol p')=E_\pi(\boldsymbol p)$, and at $t=\tau/2$ the temporal component ($\mu=0$) of the full three-point function takes the form
\begin{equation}
C_0^{AB}(\tau,\tau/2;-\boldsymbol{p},\boldsymbol{p}) = \frac{Z_\pi^A Z_\pi^B}{2E_\pi(\boldsymbol{p})} e^{-E_\pi(\boldsymbol{p})\tau}F_\pi(Q^2) + \delta C_0^{AB}(\tau,\tau/2;-\boldsymbol{p},\boldsymbol{p})\,, \label{eq:C4_back_to_back}
\end{equation}
where \(Q^2=4\boldsymbol{p}^2\), with the right-hand side understood as the analytic continuation of this expression for complex \(\boldsymbol{p}\) within the convergence domain. The excited-state contamination \(\delta C_0^{AB}\) is defined by the spectral representation in eq.~\eqref{eq:delta_C_spectral_density} and is suppressed relative to the ground-state term at large \(\tau\). We emphasize that the universal overlap factors \(Z_\pi^A\) and \(Z_\pi^B\) appear here as well, again evaluated at \(p^2 = - m_\pi^2\). In the large-\(\tau\) limit, these factors and the exponential decay can be removed by forming a ratio with the two-point function, which gives
\begin{equation}
\label{eq:ff_ratio_2pt} F_\pi(Q^2) = \lim_{\tau\to\infty} \frac{C_0^{AB}(\tau,\tau/2;-\boldsymbol{p},\boldsymbol{p})} {G^{AB}(\tau,\boldsymbol{p})}\,.
\end{equation}

Variants of this approach based on fitting before forming ratios are possible. For example, because \(Z_\pi^A\) and \(Z_\pi^B\) are independent of the spatial momentum in the infinite-volume continuum limit, it is viable to determine these quantities, together with \(m_\pi\), from the zero-momentum projected two-point function. An estimator for the ground-state dominated two-point function can then be formed using these results together with the relativistic dispersion relation to evaluate \(E_\pi(\boldsymbol{p})\) for any complex-valued momentum accessible in the numerator. While this gives the universal result in the infinite-volume continuum limit, the fact that the overlap factors and energies are determined in a fixed frame could enhance discretization and finite-volume effects. Therefore another alternative is to perform the fits to the two-point function at the same complex momenta used in the three-point function, determining \(E_\pi(\boldsymbol{p})\) and the overlap factors in a way that matches finite-volume and discretization effects between the two- and three-point functions.

A final option follows from the observation that the ground-state contribution to the three-point function is fixed when evaluated at \(\boldsymbol{p}'=\boldsymbol{p}\). Including excited-state contamination, in that case one finds
\begin{equation}
C_0^{AB}(\tau,\tau/2;\boldsymbol{p},\boldsymbol{p}) = \frac{Z_\pi^A Z_\pi^B}{2E_\pi(\boldsymbol{p})} e^{-E_\pi(\boldsymbol{p})\tau} + \delta C_0^{AB}(\tau,\tau/2;\boldsymbol{p},\boldsymbol{p})\,, \label{eq:C4_equal_momenta}
\end{equation}
where we have used \(F_\pi(0)=1\), and the final term is again defined in eq.~\eqref{eq:delta_C_spectral_density}. Since this contamination is asymptotically suppressed, another estimator is given by
\begin{equation}
\label{eq:ff_ratio_3pt} F_\pi(Q^2) = \lim_{\tau\to\infty} \frac{C_0^{AB}(\tau,\tau/2;-\boldsymbol{p},\boldsymbol{p})} {C_0^{AB}(\tau,\tau/2;\boldsymbol{p},\boldsymbol{p})}\,.
\end{equation}
This estimator has the advantage that at $Q^2=0$ the numerator and denominator are identical, so the ratio reduces exactly to $1$ for every value of $\tau$, not only in the $\tau\to\infty$ limit. Whether or not this is an advantage for determining the form factor at general $Q^2$ needs to be tested using numerical data, as we do in section~\ref{sec:numerical_check}.

For completeness, in appendix~\ref{app:position_space_moments} we also describe how the same position-space perspective can be used to obtain momentum derivatives of form factors from coordinate moments of the correlators.

Real timelike momentum transfers $Q^2 \in (-4m_\pi^2, 0)$ are accessed with purely imaginary back-to-back momenta $\vec{p} = -\vec{p}' = i\vec{\kappa}$, $\vec{\kappa} \in \mathbb{R}^3$, giving
\begin{equation}
F_\pi(- 4 \boldsymbol{\kappa}^2) = \frac{2\sqrt{m_\pi^2 - \boldsymbol{\kappa}^2}}{Z_\pi^A Z_\pi^B } \lim_{\tau \to \infty} e^{\sqrt{m_\pi^2 - \boldsymbol \kappa^2} \tau} \int \!{\rm d}^3 \boldsymbol{x} \! \int \!{\rm d}^3 \boldsymbol{y} \, e^{\boldsymbol{\kappa} \cdot (\boldsymbol{x} + \boldsymbol{y})} C^{AB}_{0}(\tau,\tau/2;\boldsymbol{x},\boldsymbol{y}) \,.
\end{equation}
The overlap factors \(Z_\pi^A\) and \(Z_\pi^B\) and the exponential factor involving the mass $m_\pi$ are assumed to be obtained by one of the methods discussed above. In words, a linear integral transform of the position-space Euclidean three-point correlator, evaluated at large Euclidean time separations and with the overlap factors removed, directly yields the form factor at timelike momentum transfer $Q^2 = - 4 \boldsymbol{\kappa}^2$.

\section{Finite-volume estimators} \label{sec:finite_volume}

In a finite periodic volume, a transform evaluated at complex or non-Fourier momentum does not project onto an energy eigenstate with that momentum. Instead, it overlaps with the discrete Fourier spectrum and is eventually dominated at large Euclidean time by lower-energy modes. This section characterizes that obstruction and considers three ways to control it: increasing or extrapolating the volume, reconstructing the missing position-space tails, and modifying the boundary conditions so that the desired complex momenta are supported directly.

Lattice-spacing effects may also be a concern in practical numerical calculations. The numerical study in section~\ref{sec:numerical_check} provides preliminary evidence that the complex-momentum transforms do not themselves lead to unusually large lattice artifacts, but this conclusion is based on a single lattice spacing and requires a dedicated study with a controlled continuum extrapolation. More generally, logarithmic corrections can modify the naive Symanzik scaling of cutoff effects~\cite{Husung:2019ytz,Husung:2022kvi}. Controlling such effects is an important component of recent lattice determinations of the hadronic-vacuum-polarization contribution to muon $g-2$~\cite{Kuberski:2024bcj,Spiegel:2024dec,Boccaletti:2024guq,ExtendedTwistedMass:2024nyi,MILC:2024ryz,Bazavov:2024eou}.

\subsection{Periodic boundary conditions}

\begin{figure}[t] \centering \includegraphics{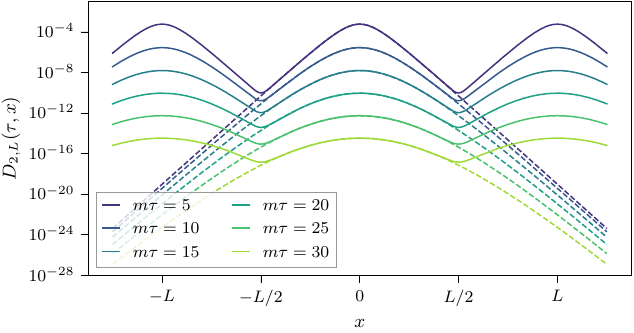} \caption{Schematic illustration of the periodic position-space propagator (solid lines) as a sum over images of the infinite-volume propagator (dashed lines). The dependence on both spatial separation $x$ and temporal separation $\tau$ is shown. For larger values of $\tau$, the finite-volume effects relative to the total magnitude of the correlator are larger. \label{fig:periodic_propagator}} \end{figure}

Most lattice QCD simulations are carried out in spatial volumes with periodic boundaries for both the quark and gluon fields. The periodicity of the quarks and gluons implies that all hadronic operators built from these fields are also periodic. For example, the pion interpolating operator satisfies
\begin{equation}
\pi^a_A(x + L \hat{\boldsymbol{e}}_i) = \pi^a_A(x).
\end{equation}
Here, $\hat{\boldsymbol{e}}_i$ denotes the unit vector in the $i$th spatial direction. As a result, the position-space correlation functions are also periodic. As a concrete example, we recall that the $3+1$-dimensional scalar two-point function in a non-interacting infinite-volume theory is
\begin{equation}
\label{eq:free-prop-4d}
D(x) = \int \frac{{\rm d}^4 p}{(2\pi)^4} \frac{e^{i p \cdot x}}{p^2 + m^2} = \frac{m}{4\pi^2 \lVert x\rVert} K_1(m \lVert x\rVert),
\end{equation}
where $K_n$ is a modified Bessel function of the second kind. Projecting two of the four dimensions to zero momentum, one has
\begin{equation}
\label{eq:free-prop-2d}
D_2(\tau, x) = \int \! {\rm d} y \! \int \! {\rm d} z \, D(x) = \frac{1}{2\pi} K_0(m \sqrt{\tau^2 + x^2}) \,.
\end{equation}
Equivalently, this is the definition of the Euclidean propagator in $1+1$ dimensions. In a finite volume with periodic boundary conditions, the position-space propagator can be expressed as a sum over images of the infinite-volume propagator:
\begin{equation}
\label{eq:free-prop-2d-L}
D_{2,L}(\tau, x) = \sum_{n=-\infty}^{\infty} D_2(\tau, x + n L) \,.
\end{equation}
An example of this construction is illustrated in figure~\ref{fig:periodic_propagator}.

\subsection{Spectral decomposition} Periodicity of the position-space correlation functions quickly leads to the central practical challenge in applying our method: states with complex momenta are never compatible with periodic boundaries in any finite volume. The theoretical issue is most clearly seen from the perspective of the transfer matrix. Below, we write $G_L(\tau, \boldsymbol{x})$ and $C_{L,\mu}(\tau, t; \boldsymbol{x}, \boldsymbol{y})$ for the two- and three-point correlators measured in the periodic spatial volume of size $L^3$ (cf.\ eqs.~\eqref{eq:Gx} and \eqref{eq:Cxy}). Any complex or non-integer Fourier transform of the position-space correlation function, truncated to finite volume, always overlaps with the zero-momentum state in the tower of interpolated states,
\begin{equation}
\label{eq:trunc-ft-G}
\begin{aligned}
G_L(\tau, \boldsymbol{p}) &= \int_L d^3\boldsymbol{x} \, e^{-i \boldsymbol{p} \cdot \boldsymbol{x}} G_L(\tau, \boldsymbol{x}) \\ &= \frac{1}{L^3} \int_L d^3\boldsymbol{x} \, e^{-i \boldsymbol{p} \cdot \boldsymbol{x}} \sum_{\boldsymbol{n}} e^{2 \pi i \boldsymbol{n} \cdot \boldsymbol{x} / L} G_L\left(\tau, \frac{2\pi\boldsymbol{n}}{L}\right) \\ &= \sum_{\boldsymbol{n}} c(\boldsymbol{p}, \boldsymbol{n}) \, G_L\left(\tau, \frac{2\pi\boldsymbol{n}}{L}\right).
\end{aligned}
\end{equation}
Here and below, $\int_L d^3\boldsymbol{x}$ denotes integration over one spatial unit cell. If the integrand is not periodic, as is the case for a complex or non-integer Fourier mode, the choice of cell matters and effectively defines a periodic continuation. We take each coordinate $x_i$ to run over $(-L/2,L/2)$. With this convention, the overlap coefficients between the complex or non-integer Fourier mode $\vec{p}$ and the available lattice Fourier modes $\vec{n}$ are
\begin{equation}
\label{eq:p-n-overlap} c(\boldsymbol{p}, \boldsymbol{n}) = \frac{1}{L^3} \int_{-L/2}^{L/2} d^3\boldsymbol{x} \, e^{-i (\boldsymbol{p}-2\pi \boldsymbol{n}/L) \cdot \boldsymbol{x}} = \prod_{i} \mathrm{sinc}(\tfrac{p_i L}{2} - \pi n_i).
\end{equation}
A similar decomposition can be performed for both momenta in the three-point function,
\begin{equation}
\begin{aligned} C_{L,\mu}(\tau, t; \boldsymbol{p}', \boldsymbol{p}) &= \int_{-L/2}^{L/2} d^3\boldsymbol{x} \, e^{-i \boldsymbol{p}' \cdot \boldsymbol{x}} \int_{-L/2}^{L/2} d^3\boldsymbol{y} \, e^{i \boldsymbol{p} \cdot \boldsymbol{y}} C_{L,\mu}(\tau, t; \boldsymbol{x}, \boldsymbol{y}) \\ &= \sum_{\boldsymbol{n}', \boldsymbol{n}} c(\boldsymbol{p}', \boldsymbol{n}') \, c(\boldsymbol{p}, \boldsymbol{n}) \, C_{L,\mu}\left(\tau, t; \frac{2\pi\boldsymbol{n}'}{L}, \frac{2\pi\boldsymbol{n}}{L} \right), \end{aligned}
\end{equation}
which relates complex or non-integer Fourier modes $\boldsymbol{p}',\boldsymbol{p}$ to lattice momenta $\vec{n}', \vec{n}$ via a product of the same overlap factors. From eq.~\eqref{eq:p-n-overlap}, the overlap onto $\boldsymbol{n} = 0$ vanishes only if at least one component of $\vec{p}$ is an integer multiple of the finite-volume momentum $2\pi/L$. The zero-momentum state is therefore guaranteed to dominate at large Euclidean time separation for generic complex or non-integer Fourier modes.

This seems to imply that one cannot extrapolate to infinite Euclidean time to isolate the overlap factor or form factor contributions associated with the correct physical state. On the other hand, we expect that performing truncated Fourier transforms of position-space data with sufficiently suppressed tail contributions should give results that are identical to the infinite-volume setting, up to exponentially small corrections. This tension can be reconciled by noting that the order of limits matters. For this exponential convergence, the spatial volume must be sufficiently large for each fixed value of $\tau$, and for any finite $L$ this statement does not hold for $\tau \to \infty$. Care is therefore needed to extract the relevant physical quantities. We propose two solutions:
\begin{enumerate}
\item Enforce an appropriate order of limits, first evaluating the $L \to \infty$ limit for each Euclidean time $\tau$, then extrapolating $\tau \to \infty$ (e.g.\ by identifying an appropriate plateau or fitting the data). Section~\ref{subsec:extrapolating} discusses this approach.
\item Change the transfer matrix by modifying the boundary conditions of the theory, so that states with the desired momenta are included in the spectrum directly. Section~\ref{subsec:bcs} discusses this option.
\end{enumerate}

\begin{figure} \centering \includegraphics{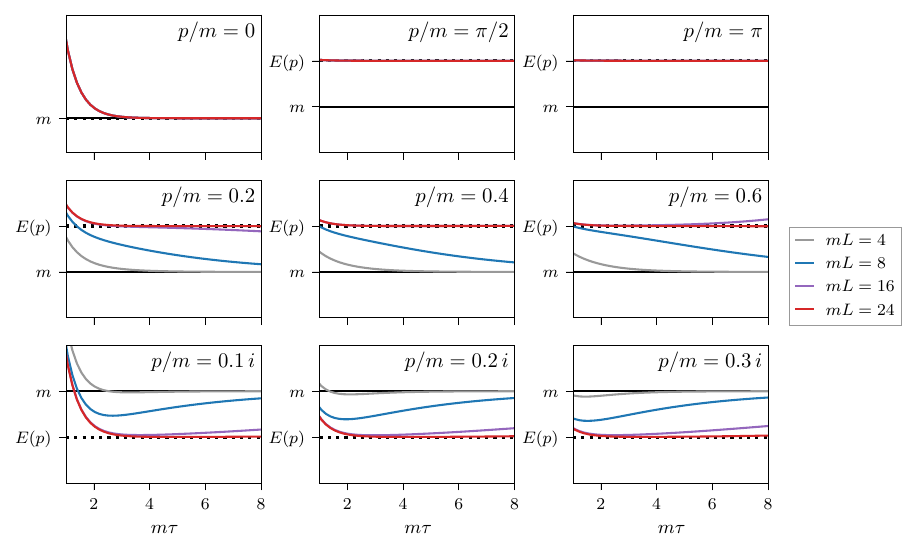} \caption{Comparison of effective masses, $m_{L,{\rm eff}}(\tau,p) = -\partial_\tau \ln D_{2,L}(\tau,p)$, extracted from the position-space correlator of the non-interacting scalar theory using direct finite-volume (complex) Fourier transforms. A variety of momenta are shown, including complex and non-integer multiples of $2\pi/L$. As $\tau \to \infty$, all effective masses shown converge to the ground state $m$. On the other hand, taking the infinite-volume limit $L \to \infty$ for fixed separation $\tau$, each trace converges to the analytically continued energy $E(\boldsymbol{p})$ up to excited-state contributions.} \label{fig:meff-iv-limit} \end{figure}

Figure~\ref{fig:meff-iv-limit} shows a representative example of this issue and its resolution for the non-interacting scalar theory. Each panel depicts the effective mass of the Fourier-transformed finite-volume correlator for various choices of momentum $p$ in the $\boldsymbol{e}_3$ direction. Using the dimensionally reduced propagator defined in the previous section, the relevant effective mass can be written as
\begin{equation}
m_{L,{\rm eff}}(\tau,p) = -\partial_\tau \log D_{2,L}(\tau, p), \quad D_{2,L}(\tau, p) \equiv \int_{-L/2}^{L/2} dx_3 \, e^{i p x_3} D_{2,L}(\tau, x_3),
\end{equation}
where the symmetric range $[-L/2, L/2]$ is used to define the truncated non-integer or complex Fourier transforms as in eq.~\eqref{eq:trunc-ft-G}. For finite-volume-quantized momenta $p = 2\pi n / L$ shown in the top row of panels, the effective mass converges to the expected energy given by the dispersion relation, $E(p) = \sqrt{p^2 + m^2}$, whereas the non-integer and complex momenta shown in the bottom two panels converge towards the ground state energy $m$ as $\tau \to \infty$. However, the estimated effective mass for \emph{each value of $\tau$} converges to the expected energy given by the dispersion relation as the infinite volume limit $L \to \infty$ is taken, demonstrating that an appropriate order of limits can resolve the issue.

\subsection{Extrapolating to infinite spatial volume} \label{subsec:extrapolating} As the spatial volume is increased, the difference between the finite- and infinite-volume correlation functions decreases exponentially. The absolute error can be bounded by summing the effects of the ``missing tails'' outside the finite volume and the ``wrap-around'' (and self-interacting) contributions to the correlator within the finite volume,
\begin{equation}
\abs{G_{\infty}(\tau, \boldsymbol{p}) - G_L(\tau, \boldsymbol{p})} \leq \Delta_{\rm tail} + \Delta_{\rm wrap},
\end{equation}
where
\begin{align}
\Delta_{\rm tail} &\equiv \int_{\abs{x_i} \geq L/2} d^3\boldsymbol{x} \, e^{\Im{\boldsymbol{p}} \cdot \boldsymbol{x}} \abs{G_{\infty}(\tau, \boldsymbol{x})}, \\ \Delta_{\rm wrap} &\equiv \int_{-L/2}^{L/2} d^3{\boldsymbol{x}} \, e^{\Im{\boldsymbol{p}} \cdot \boldsymbol{x}}\abs{G_{\infty}(\tau,\boldsymbol{x}) - G_L(\tau, \boldsymbol{x})}.
\end{align}
Due to the mass gap of the theory, $|G_{\infty}(\tau,\boldsymbol{x})|$ falls exponentially as $e^{-m \lVert\boldsymbol{x}\rVert}$, up to powers of $\lVert\boldsymbol{x}\rVert$, at fixed $\tau$ and sufficiently large spatial distances. Consequently, for $\lVert \Im\vec{p} \rVert < m$, both finite-volume contributions satisfy the conservative asymptotic bound
\begin{equation}
\Delta_{\rm tail},\,\Delta_{\rm wrap} = O\!\left(L^k e^{-\frac{1}{2}(m-\lVert \Im\vec{p} \rVert)L}\right),
\end{equation}
where $k$ denotes an unknown power determined by the power-law prefactor in the large-distance correlator and by the geometry of the integration region. This bound need not be saturated, since cancellations between missing-tail and wrap-around contributions can improve the convergence for special momenta. In practice, the reduced exponential suppression makes $\Delta_{\rm tail}$ particularly sensitive to growing or slowly falling exponentials when $\lVert \Im\vec{p} \rVert \approx m$. Together, these bounds guarantee that the simple finite-volume estimator $G_L(\tau, \boldsymbol{p})$ converges exponentially to the infinite-volume result as $L \to \infty$, for fixed $\tau$ and for all $\boldsymbol{p}$ within the region of analyticity.

For the dimensionally reduced non-interacting example in eqs.~\eqref{eq:free-prop-4d}--\eqref{eq:free-prop-2d-L}, the wrap-around contribution can be made explicit. Setting $\kappa\equiv\Im p$ and using the positivity of $K_0$ gives
\begin{equation}
\Delta_{\rm wrap}=\frac{1}{2\pi}\sum_{n\ne0}\int_{-L/2}^{L/2}{\rm d}x\,e^{\kappa x}K_0\!\left(m\sqrt{\tau^2+(x+nL)^2}\right).
\end{equation}
At fixed $\tau$ and large $L$, the nearest images dominate. Using $K_0(mr)\sim\sqrt{\pi/(2mr)}e^{-mr}$, one obtains
\begin{equation}
\begin{aligned}
\Delta_{\rm wrap}&\sim\frac{1}{2\pi}\sqrt{\frac{\pi}{mL}}e^{-mL/2}\left[\frac{e^{\kappa L/2}}{m+\kappa}+\frac{e^{-\kappa L/2}}{m-\kappa}\right] \\
&\sim\frac{1}{2\pi(m+|\kappa|)}\sqrt{\frac{\pi}{mL}}e^{-\frac{1}{2}(m-|\kappa|)L},
\end{aligned}
\end{equation}
where the final expression holds for fixed nonzero $\kappa$ with $|\kappa|<m$.
The missing-tail contribution follows directly by integrating the same large-distance form over $|x|\geq L/2$:
\begin{equation}
\begin{aligned}
\Delta_{\rm tail}&\sim\frac{1}{2\pi}\sqrt{\frac{\pi}{mL}}e^{-mL/2}\left[\frac{e^{\kappa L/2}}{m-\kappa}+\frac{e^{-\kappa L/2}}{m+\kappa}\right] \\
&\sim\frac{1}{2\pi(m-|\kappa|)}\sqrt{\frac{\pi}{mL}}e^{-\frac{1}{2}(m-|\kappa|)L},
\end{aligned}
\end{equation}
where the second line again holds for fixed nonzero $\kappa$ with $|\kappa|<m$. Thus the unknown power in the general bound is $k=-1/2$ for this example. Compared with $\Delta_{\rm wrap}$, the coefficient multiplying the leading exponential in $\Delta_{\rm tail}$ is enhanced by $1/(m-|\kappa|)$ rather than $1/(m+|\kappa|)$, making the missing-tail contribution parametrically more sensitive as $|\kappa|\to m$. At $\kappa=0$, both terms separately scale as $\sqrt{\pi/(mL)}e^{-mL/2}/(\pi m)$, but they cancel exactly in the difference between the infinite-volume transform and the transformed image sum. This explicitly illustrates why the separate positive bounds can overestimate the actual finite-volume error for special momenta.

A simple strategy to extract overlap factors $Z_A$, $Z_B$, energies $E(\vec{p})$, and form factors $F(Q^2)$ is thus to work with a sufficiently large volume to control such corrections. Unfortunately, initial evidence from figure~\ref{fig:meff-iv-limit} and numerical evidence explored in sec.~\ref{sec:numerical_check} below suggest that the values of $mL$ required to achieve acceptable control of errors are larger than the typical value $m_\pi L \sim 4$ used in current lattice QCD calculations, at least for the specific non-interacting and two-dimensional examples considered here. These thresholds will in general differ for lattice QCD correlators, but this motivates exploring other ways to control the infinite-volume limit.

A more robust alternative is to perform an asymptotic analysis of the tail region of the correlator. For example, a spectral decomposition in the spatial direction is possible, and analogous ideas have been used successfully in the analysis of the hadronic vacuum polarization contribution to the muon $g-2$~\cite{DellaMorte:2017khn,RBC:2024fic,Boccaletti:2024guq,Bazavov:2024eou,Lehner:2025qrl}. Because operators are not constructed to project out particular momenta of the spatial transfer matrix, it may be necessary to reconstruct many such states to control this approach. Alternatively, combined position-space fits in all Euclidean directions may be used. By systematically extracting the leading and sub-leading exponential behavior of $G_{\infty}(\tau, \boldsymbol{x})$ as $\lVert\boldsymbol{x}\rVert \to \infty$, the Fourier transform can be extended to mitigate the truncation, reducing or removing $\Delta_{\rm tail}$.
The remaining finite-volume effects $\Delta_{\rm wrap}$ in the bulk are no different from those observed in standard lattice calculations. Controlling them with large physical volumes, extrapolating the final physical quantities, or both should suffice to obtain reliable results from the method in this work.

\subsection{Spatial chemical potential} \label{subsec:bcs} A second strategy to solve the finite-volume issue is to calculate the correlation functions in a version of the theory with a transfer matrix that directly supports the momentum states of interest. For example, continuous real momenta can be accessed in a finite volume using twisted boundary conditions~\cite{Bedaque:2004kc,Sachrajda:2004mi}. We begin by reviewing this approach, then discuss its generalization to complex momenta.

Twisted boundary conditions are typically imposed on the quark fields in lattice QCD. We assume a cubic finite volume with extent $L$ in each direction. For each spatial direction $\hat{\boldsymbol{e}}_i$ ($i \in \{1,2,3\}$), impose
\begin{equation}
\label{eq:twisted-bcs} q(\tau,\boldsymbol{x} + L \hat{\boldsymbol{e}}_i) = e^{i \Theta_{i}} q(\tau,\boldsymbol{x}), \qquad \bar{q}(\tau,\boldsymbol{x} + L \hat{\boldsymbol{e}}_i) = e^{-i \Theta_{i}} \bar{q}(\tau,\boldsymbol{x}),
\end{equation}
where $q$ and $\bar{q}$ respectively indicate the quark and antiquark fields, and the set of twist matrices $\Theta_i$ can be flavor-diagonal or flavor-rotating. The gauge fields are typically kept periodic, satisfying the ordinary boundary conditions
\begin{equation}
U_\mu(\tau,\boldsymbol{x} + L \hat{\boldsymbol{e}}_i) = U_\mu(\tau,\boldsymbol{x}).
\end{equation}

If the twist matrices take the form $\Theta_i=\theta_i T$, where $T$ is the generator of a conserved $U(1)$ charge, the effect of the boundary conditions is particularly simple to analyze. In this case, the finite-volume momenta depend on both the lattice extent $L$ and the charge $Q$ of a state:
\begin{equation}
\label{eq:twisted-mom} \boldsymbol{p} = \frac{2 \pi \boldsymbol{n}}{L} + \frac{Q \boldsymbol{\theta}}{L}, \qquad \boldsymbol{n} \in \mathbb{Z}^3,
\end{equation}
in terms of the twist angles $\boldsymbol{\theta} \equiv (\theta_1, \theta_2, \theta_3)$. The set of available momenta can then be continuously varied by tuning $\vec{\theta}$.

From eq.~\eqref{eq:twisted-mom}, it is apparent that one can access complex momenta by promoting the twists $\theta_1, \theta_2, \theta_3$ to complex values, $\boldsymbol{\theta} \in \mathbb{C}^3$. In the simplest case of flavor-independent complex twists, where $T = \mathrm{Id}$, the twists act on all states with nonzero baryon number, giving all such states complex momenta. For the prototypical example of the $I = 1$ pion form factor considered in this work, a natural choice of $U(1)$ charge would instead be the $z$ component of isospin, $I_z$. Three-point correlation functions using the $I_z = \pm 1$ states $\pi_{\pm}$ have momenta with contributions $\pm \boldsymbol{\theta}$, which can continuously access the complex plane.

Complexified twisted boundaries amount to combining ordinary twisted boundary conditions with a chemical potential applied in each of the three spatial dimensions. This can be seen directly from the form of the twist in eq.~\eqref{eq:twisted-bcs}, which mirrors an applied chemical potential for imaginary $\theta_i$. In general, introducing such chemical potentials creates a sign problem that can obstruct standard Monte Carlo estimates. Fortunately, isospin chemical potential has no sign problem in QCD, because the phases from the up and down quarks cancel~\cite{Kogut:2002tm,Kogut:2002zg}, making this a viable method for the pion form factor in lattice QCD. We leave numerical investigation of this strategy and its generalizations to future work.

\section{Numerical check in the O(3) NLSM} \label{sec:numerical_check}

To demonstrate the proposed method, we perform numerical calculations of the $I=1$ form factor $F(Q^2)$ for timelike and complex momentum transfer in the two-dimensional $O(3)$ nonlinear sigma model (NLSM). This model has a mass gap $m$, and its lightest particles form an $I=1$ representation of the $\mathfrak{so}(3) \simeq \mathfrak{su}(2)$ isospin symmetry, in close analogy with the pion sector of QCD. We evaluate the form factor from Monte Carlo data using methods inspired by those of ref.~\cite{Luscher:1990ck}.

\begin{figure} \centering \includegraphics{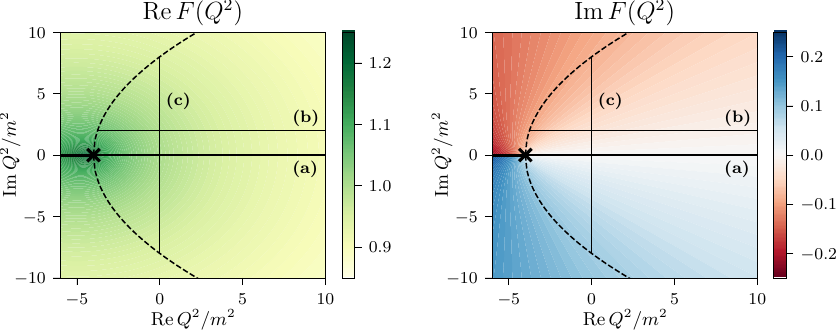} \caption{Real and imaginary parts of $F(Q^2)$ on the first sheet in the $Q^2$ plane. The bold line (cross) on the left side of each panel indicates the particle-production branch cut (branch point). The dashed parabola indicates the boundary of the accessible region with back-to-back complex momenta. Lines (a)--(c) indicate specific kinematics investigated numerically below.} \label{fig:ff_grid_thy} \end{figure}

Because the theory is integrable, the form factor can be evaluated analytically throughout the complex plane, giving~\cite{Zamolodchikov:1978xm}
\begin{equation}
\label{eq:o3-ff} F(Q^2) = \frac{\theta}{2 \tanh(\theta/2)} \frac{\pi^2}{\pi^2 + \theta^2}, \qquad \theta = \arccosh\left(1 + \frac{Q^2}{2m^2}\right).
\end{equation}
Taking the principal value of arccosh gives $\Re{\theta} \in [0,\infty)$ and $\Im{\theta} \in [-\pi,\pi]$, which places the form factor on the first sheet, where the branch cut is placed along the line of real particle production, $Q^2 \in (-\infty, -4m^2)$. Figure~\ref{fig:ff_grid_thy} shows these first-sheet results, and below we compare eq.~\eqref{eq:o3-ff} against our numerical data.

\subsection{Basic setup} Throughout, we work with the simple lattice discretization of the $O(3)$ NLSM given by
\begin{equation}
S = -\beta \sum_{t, x} [\sigma(t, x) \cdot \sigma(t, x+1) + \sigma(t, x) \cdot \sigma(t+1, x)],
\end{equation}
where $\sigma(t,x) \in \mathbb{R}^3$ with $\lVert \sigma(t,x) \rVert = 1$. The observables needed to extract the form factor are the two-point and three-point functions, defined respectively by
\begin{align}
G(\tau,x) &\equiv \frac{1}{3} \delta_{ab} \langle \sigma_a(\tau,x) \, \sigma_b(0,0) \rangle, \\ C_0(\tau,t; x,y) &\equiv \frac{1}{6} \epsilon_{abc} \langle \sigma_a(\tau,x) \, j^0_{b}(t,0) \, \sigma_c(0,y) \rangle,
\end{align}
where the discretized $I=1$, $\mu=0$ current is defined by
\begin{equation}
j^0_a(t,x) \equiv \beta \, \epsilon_{abc} \, \sigma_b(t,x) \, \sigma_c(t+1,x).
\end{equation}

The two- and three-point correlation functions are estimated using Monte Carlo simulation with a double-cluster algorithm that enables improved observables for both kinds of correlators~\cite{Wolff:1988uh,Bulava:2021fre}. Each double-cluster update randomly samples two orthogonal axes $\hat{\boldsymbol{e}}_1$ and $\hat{\boldsymbol{e}}_2$ in spin space, generating independent clusters according to the standard prescription for each reflection axis. Within each update, we define the longitudinal projections $\ell_1(x) \equiv \sigma(x) \cdot \hat{\boldsymbol{e}}_1$ and $\ell_2(x) \equiv \sigma(x) \cdot \hat{\boldsymbol{e}}_2$ as well as indicator functions $I_{1,2}(t,x; s,y)$ that equal $1$ if the spins at sites $(t,x)$ and $(s,y)$ belong to the same first- or second-axis cluster, respectively, and $0$ otherwise. An improved estimator of the two-point function can then be written in terms of either cluster projection,
\begin{equation}
G(\tau,x) = \left< I_1(\tau,x; 0,0) \ell_1(\tau,x) \ell_1(0,0) \right> = \left< I_2(\tau,x; 0,0) \ell_2(\tau,x) \ell_2(0,0) \right>.
\end{equation}
Below, we use the first cluster axis $\hat{\boldsymbol{e}}_1$ for all estimates of $G(\tau,x)$. An improved estimator of the three-point function can similarly be written, using both projection axes, as
\begin{equation}
\begin{aligned} C_0(\tau,t; x,y) &= \beta \, \big\langle I_1(\tau,x; t+1,0) \ell_1(\tau,x) \ell_1(t+1,0) \, I_2(t,0; 0,y) \ell_2(t,0) \ell_2(0,y) \\ &\qquad- I_1(\tau,x; t,0) \ell_1(\tau,x) \ell_1(t,0) \, I_2(t+1,0; 0,y) \ell_2(t+1,0) \ell_2(0,y) \big\rangle. \end{aligned}
\end{equation}

All results below are based on two Monte Carlo ensembles with spatial extents $L = 128$ and $L = 256$ and a common temporal extent $T = 64$. Our main results for the form factor use the larger spatial volume, while the smaller volume provides useful checks of finite-volume effects. Each ensemble is generated with $5 \times 10^6$ cluster updates, and two- and three-point functions are measured using $10$ random sources per configuration. The bare coupling is fixed to $\beta = 1.4$, corresponding to a correlation length $\xi \approx 7$ in lattice units.

\subsection{Form factor calculations} To extract the form factor from the correlation functions above, we restrict attention to the back-to-back momentum frame. In infinite volume, the relevant projection of the three-point function is then
\begin{equation}
C_0(\tau, t; -p, p) = \int dx \, dy \, e^{i p (x + y)} C_0(\tau, t; x, y),
\end{equation}
where in general $p \in \mathbb{C}$. Following the identities in eqs.~\eqref{eq:ff_ratio_2pt} and \eqref{eq:ff_ratio_3pt}, we explore two kinds of ratios that cancel the overlap factor $Z$ and the exponential factors. We consider both the ratio with the two-point function~\cite{Luscher:1990ck},
\begin{equation}
R(Q^2; t) \equiv \frac{C_0(\tau, t; -p, p)}{G(\tau, p)},
\end{equation}
and the ratio with the zero-momentum-transfer three-point function,
\begin{equation}
\bar{R}(Q^2; t) \equiv \frac{C_0(\tau, t; -p, p)}{C_0(\tau, t; p, p)},
\end{equation}
where $\tau = 2t+1$. In both cases the virtuality is set by the complex Fourier transform according to $Q^2 = 4 p^2$. The form factor is then obtained from the limit
\begin{equation}
F(Q^2) = \lim_{t \to \infty} R(Q^2; t) = \lim_{t \to \infty} \bar{R}(Q^2; t).
\end{equation}

Both ratios are written formally in terms of the infinite-volume time-momentum correlation functions. As discussed in sec.~\ref{sec:finite_volume}, several strategies can connect the position-space correlation functions that we measure in finite volume to the theoretical infinite-volume ratios. The simplest option is to perform the truncated Fourier transform directly, integrating between $-L/2$ and $L/2$ in each relevant coordinate. As explored below, this yields poor results even for the larger volume studied here, motivating a comparison with more robust methods.

In contrast, our main results for the numerators of $R(Q^2; t)$ and $\bar{R}(Q^2; t)$ are based on fitting the tails of the three-point function and replacing the data with an infinite-volume extension of the fit form before applying the complex Fourier transform. To simplify the fits, we first integrate one spatial component,
\begin{equation}
C(t, w) \equiv \int dx \, C_0(\tau, t; x, w-x),
\end{equation}
which leaves only the dependence on $w = x+y$. This projection is then fit to the simple functional form
\begin{equation}
f(t, w) = A \exp(-E\sqrt{w^2 + t^2}),
\end{equation}
where $A$ and $E$ are fit parameters. In this exploratory study, all fits minimize chi-squared using $C(t,w)$ data for $t \in [2, 11]$ and $|w| \geq 50$. For the final time-momentum results, the tails of the projection $C(t,w)$ are replaced with the fit function for all $|w| \geq x_{\rm cut}$ before applying the complex Fourier transform according to
\begin{equation} \label{eq:three-pt-data-plus-fit}
C_0(\tau,t; -p, p) = \int_{-x_{\rm cut}}^{x_{\rm cut}} dw \, e^{i p w} \, C(t, w) + 2 \int_{x_{\rm cut}}^{\infty} dw \, \cos(p w) f(t, w).
\end{equation}
For $L = 256$ data, we explore $x_{\rm cut} \in \{40, 80, 120\}$, while for $L = 128$ only $x_{\rm cut} = 40$ is valid. For the denominator of $\bar{R}(Q^2; t)$, we apply an identical procedure using the three-point function integrated in the orthogonal direction,
\begin{equation}
C'(t, w) \equiv \int dx \, C_0(\tau, t; x, w+x),
\end{equation}
from which a fit and Fourier transform analogous to eq.~\eqref{eq:three-pt-data-plus-fit} yield the denominator factor $C_0(\tau, t; p, p)$.

\begin{figure} \centering \includegraphics[width=0.88\textwidth]{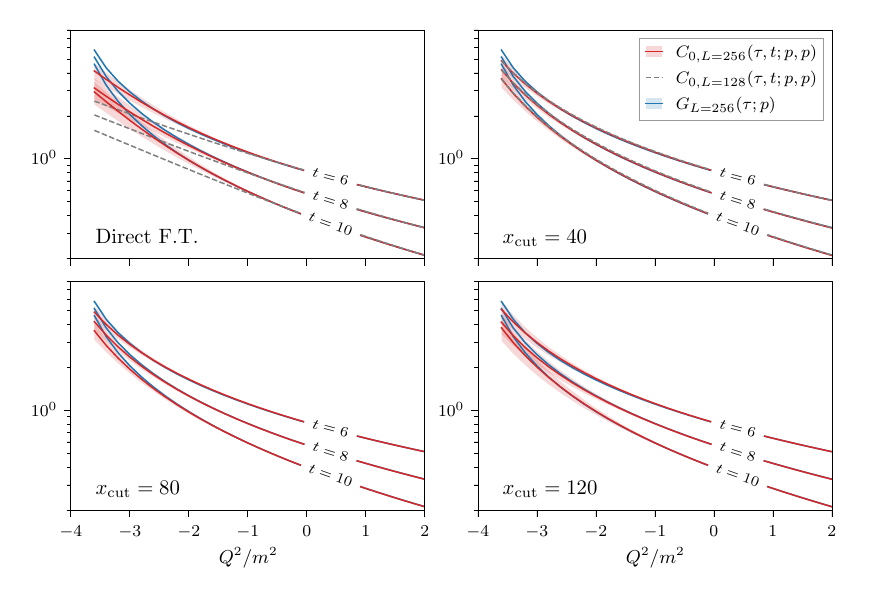} \caption{Comparison of strategies for estimating the two-point function that enters the ratios used to extract the form factor. In each case, estimates using both the zero-momentum-transfer three-point function and two-point function are compared. As discussed in the main text, results for the two-point function are obtained by fitting and analytically continuing energies, while for the three-point function they are obtained either by direct Fourier transform (top-left panel) or by fitting and replacing tails beyond $x_{\rm cut} \in \{40, 80, 120\}$ by the infinite-volume fit function (other panels). For clarity, only the central values are shown for the $L=128$ data (dashed gray line).} \label{fig:fv_strategies} \includegraphics[width=0.88\textwidth]{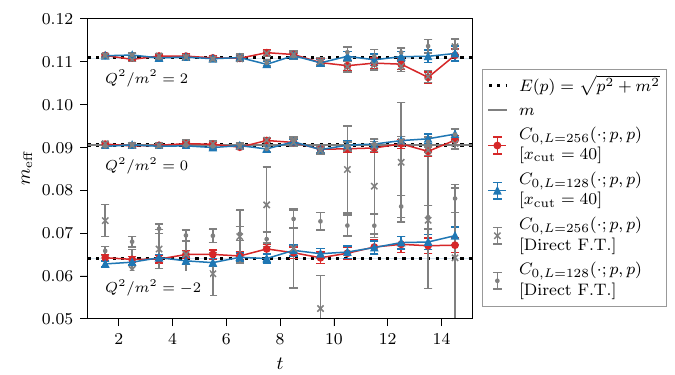} \caption{Comparison of measured effective mass versus separation $t$ using various finite-volume strategies to estimate the two-point function, which forms the denominator in the ratios used to extract the form factor. For fits to the three-point function, the value $x_{\rm cut} = 40$ is used to enable comparison of data between the $L = 128$ and $L = 256$ ensembles used in this work.} \label{fig:ff_denominator_meff} \end{figure}

A similar position-space fit would be possible for the denominator of $R(Q^2; t)$; however, the Ward identity applied to the zero-momentum-transfer current exactly relates this quantity to $C_0(\tau, t; p, p)$~\cite{Luscher:1990ck}. Alternatively, the known structure of the two-point function enables another strategy, motivated by the continuum dispersion relation: fit the zero-momentum representation to obtain the overlap factor $Z_A Z_B$ and the mass $m$, then analytically continue to arbitrary $E(p) = \sqrt{m^2 + p^2}$. We explore this option to contrast it with the numerically estimated Fourier-transformed denominator.

Figure~\ref{fig:fv_strategies} shows a comparison of these strategies as they apply to estimates of the denominators of $R(Q^2; t)$ and $\bar{R}(Q^2; t)$. The top-left panel shows significant finite-volume effects in the direct truncated Fourier transform for imaginary momenta, evident both from the difference between values measured at $L=128$ and $L=256$ and from deviations from the result obtained by fitting the two-point function and applying the infinite-volume dispersion relation. In contrast, fitting the position-space three-point function and replacing the tails with the infinite-volume fit function yields consistent values across both the $L=128$ and $L=256$ data and is stable as the tail cut position is varied. Within statistical errors, the results agree well with the dispersion relation from the two-point function fit, even close to the particle production threshold at $Q^2 = -4m^2$. We also compare these strategies in extracting the effective mass of states with various momenta in figure~\ref{fig:ff_denominator_meff}. The figure compares energy levels expected from the continuum dispersion relation with numerical effective-mass estimates over a range of $\tau$ values. With the position-space fits, the results for non-integer and complex momenta remain consistent with plateaus for both $L=128$ and $L=256$. The direct truncated Fourier transform instead converges towards the ground-state energy for complex momentum (cf.\ figure~\ref{fig:meff-iv-limit}), indicating stronger finite-volume effects. These observations emphasize the importance of accounting for missing tail contributions, in this case through position-space fits.

\begin{figure}[p] \centering \includegraphics{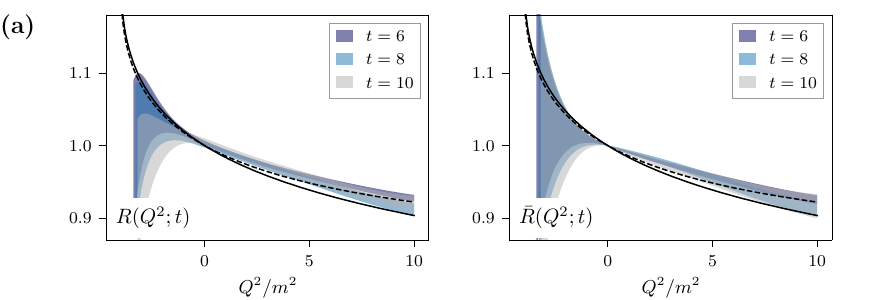} \vspace{0.1cm}

\includegraphics{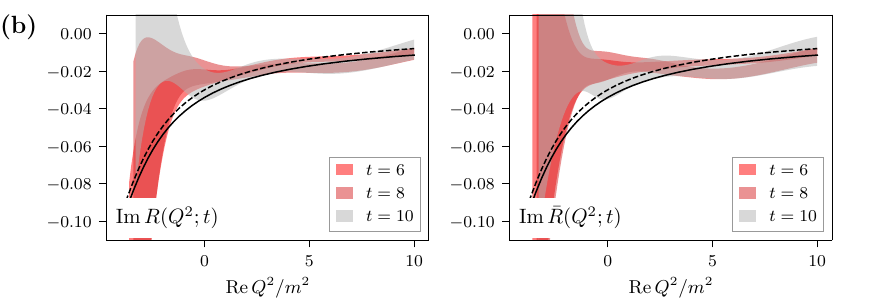} \vspace{0.1cm}

\includegraphics{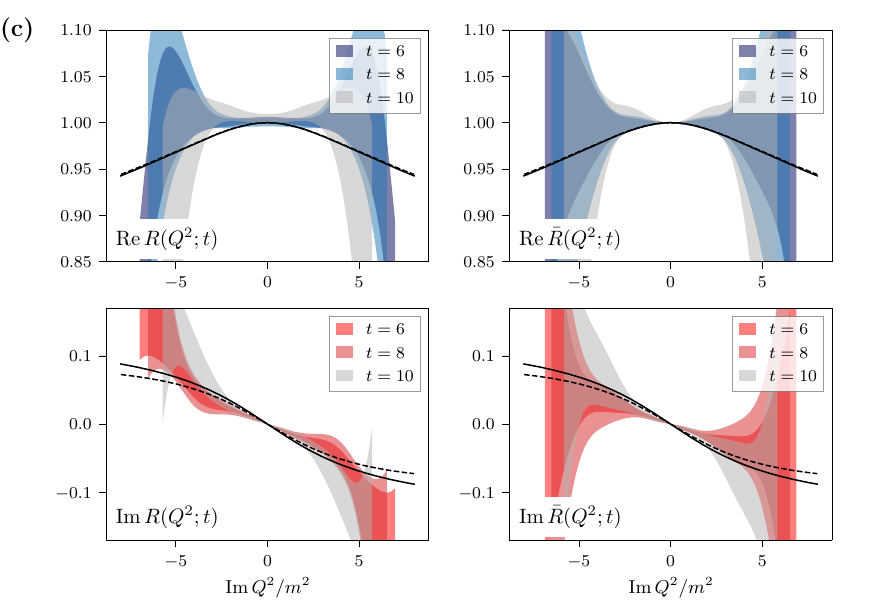}

\caption{$I=1$ form factor evaluated for (a) real $Q^2$, (b) fixed imaginary part $\Im{Q^2} = 2m^2$, and (c) imaginary $Q^2$. See figure~\ref{fig:ff_grid_thy} for a visualization of these kinematics in the $Q^2$ plane.} \label{fig:ff_data_slices} \vspace{-0.5cm} \end{figure}

\begin{figure} \centering \includegraphics{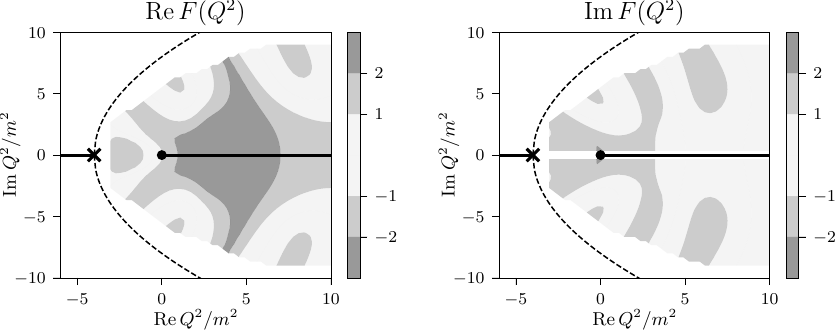} \caption{Pull in standard deviations for a numerical evaluation of the $I=1$ form factor in the region of the complex plane that is accessible via our method. Results are for the ratio $R(Q^2; t=10)$, and only points with a data fraction above 75\% are shown. The bold lines show the timelike (left) and spacelike (right) kinematics that are accessible via standard Euclidean methods.} \label{fig:ff_grid} \end{figure}

We next investigate concrete determinations of the $I=1$ form factor from the $L=256$ data, using the position-space fit procedure. Comparing determinations based on both $R(Q^2; t)$ and $\bar{R}(Q^2; t)$ provides a possible window into lattice-spacing effects and the viability of normalizing ratios based on the zero-momentum-transfer three-point function. We show only measurements for which the data contribution (the first term in eq.~\eqref{eq:three-pt-data-plus-fit}) exceeds $75\%$ of the total. Ratio data are presented for three representative Euclidean time separations $t \in \{6, 8, 10\}$; in this exploratory work, we do not fit and extrapolate the value of the ratio to $t \to \infty$, nor do we attempt to perform a continuum extrapolation. Figure~\ref{fig:ff_data_slices} shows our measurements along three slices of the complex $Q^2$ plane, corresponding to the labeled lines in figure~\ref{fig:ff_grid_thy}. In all cases, both the continuum theoretical result and an ansatz for the $O(a^2)$ correction motivated by ref.~\cite{Luscher:1990ck} are shown for comparison. Despite the presence of unquantified lattice artifacts and finite-$t$ effects, we find consistency within approximately $2\sigma$ for all data. Finally, figure~\ref{fig:ff_grid} shows measurements of $R(Q^2; t=10)$ on a grid in the complex $Q^2$ plane. The form factor can be accessed over a much broader region than is typically available with traditional Euclidean methods, while remaining consistent with the theoretical continuum form factor.

\section{Conclusions and outlook} \label{sec:conclusions}

In this work, we have shown that position-space Euclidean correlation functions provide direct access to hadronic form factors at complex momentum transfer. The central observation is that the exponential decay of the correlators guarantees convergence of their Fourier transforms within a finite strip of complex spatial momentum. Within this region, the usual ground-state saturation argument carries over by analytic continuation, allowing the form factor to be isolated without analytically continuing the Euclidean-time dependence and without solving an inverse problem. For the pion vector form factor, back-to-back momenta cover a parabolic domain with vertex at the two-pion branch point, while general kinematics cover the half-plane $\Re Q^2>-4m_\pi^2$. In particular, purely imaginary spatial momenta give direct access to the subthreshold timelike region.

This construction connects the regions traditionally studied with spacelike lattice form-factor calculations and with timelike reconstruction methods. Besides providing access to subthreshold timelike kinematics, it generates form-factor data in a continuous complex neighborhood of $Q^2=0$. Such data may improve determinations of charge radii and other derivatives at the origin, either directly through position-space moments or indirectly by strengthening fits based on analyticity, including conformal $z$-expansions. More generally, sampling the form factor away from the real axis may provide additional constraints on dispersive representations and on parametrizations designed to connect spacelike and timelike information.

The principal practical challenge is the treatment of the finite spatial volume. A truncated complex-momentum transform of a periodic correlator does not project onto a finite-volume energy eigenstate with the desired momentum and, at sufficiently large Euclidean time, is instead contaminated by lower-energy Fourier modes. Nevertheless, at fixed Euclidean separation it converges exponentially to the desired infinite-volume quantity as the spatial volume is increased. We have discussed several strategies for improving this convergence, including fits and spectral reconstructions of the missing position-space tails, and the use of a spatial chemical potential implemented through complex twisted boundary conditions.

Our numerical study of the $1+1$-dimensional $O(3)$ nonlinear sigma model provides an initial test of these ideas. Directly truncating the position-space transform leads to sizeable finite-volume effects, whereas fitting the long-distance tails and extending them beyond the measured volume gives results that are stable under changes in the volume and the matching point.

A natural next step is to test the method in lattice QCD, beginning with the pion form factor. Such studies should quantify the competing effects of statistical noise, finite Euclidean separation, lattice artifacts, and the modeling or reconstruction of position-space tails. It will also be important to compare periodic-volume estimators directly with calculations using a spatial chemical potential. The main difficulty here is that the imaginary twist must be applied to both the valence and sea quarks, a requirement that has limited analogous techniques in the past. Flow-based or out-of-equilibrium methods for transforming boundary conditions~\cite{Bonanno:2023ier,Bonanno:2024udh,Abbott:2024mix,Vadacchino:2024lob,Bonanno:2025pdp,Bonanno:2026tle} may provide a way around this issue.

Beyond the pion form factor, the same framework can be applied to other hadronic form factors and current insertions, as well as to momentum derivatives evaluated at general complex momenta. Finally, the accessible domain may be enlarged by explicitly subtracting the lowest states governing the asymptotic spatial falloff before applying the transform. Pursuing these directions will determine whether complex-momentum correlation functions can become a practical complement to established lattice methods for mapping the analytic structure of hadronic observables.

\section*{Acknowledgments}

We thank John Bulava, Xu Feng, Raoul Hodgson, Christoph Lehner, Antonin Portelli, and Fernando Romero-L\'opez for useful discussions. We also thank the Edinburgh Lattice Group Journal Club for the initial discussions that led to this work. M.T.H.~and G.K.~are supported in part by UK STFC grants ST/X000494/1 and ST/T000600/1. M.T.H. is further supported by UKRI Future Leaders Fellowship MR/T019956/1. G.K. was further supported by the European Union (ERC, FoundLatt, 101212401). Views and opinions expressed are, however, those of the authors only and do not necessarily reflect those of the European Union or the European Research Council. Neither the European Union nor the granting authority can be held responsible for them.

\appendix

\section{Position-space moments for momentum derivatives} \label{app:position_space_moments}

In this appendix we collect the relation between momentum derivatives of form factors and position-space moments of correlators, following the method of ref.~\cite{Feng:2019geu}.

The basic idea is that momentum derivatives of correlation functions can be obtained by multiplying the position-space correlators by factors of the spatial coordinates before performing the Fourier transform. Taking the complex-momentum transform with \(\boldsymbol{p}\in\mathbb C^3\) but with $\boldsymbol p' = - \boldsymbol p$, one can apply either the radial derivative \(\boldsymbol{p}\cdot\nabla_{\boldsymbol{p}}\) or the Laplacian \(\nabla^2_{\boldsymbol{p}}\). These give
\begin{align}
\boldsymbol{p}\cdot\nabla_{\boldsymbol{p}} C^{AB}_{0}(\tau,\tau/2;-\boldsymbol{p},\boldsymbol{p}) &= -i \int \!{\rm d}^3 \boldsymbol{x} \! \int \!{\rm d}^3 \boldsymbol{y} \, \boldsymbol{p}\cdot(\boldsymbol{x} + \boldsymbol{y})\, e^{-i\boldsymbol{p} \cdot (\boldsymbol{x} + \boldsymbol{y})} C^{AB}_{0}(\tau,\tau/2;\boldsymbol{x},\boldsymbol{y}) \, , \\ -\nabla^2_{\boldsymbol{p}} C^{AB}_{0}(\tau,\tau/2;-\boldsymbol{p},\boldsymbol{p}) &= \int \!{\rm d}^3 \boldsymbol{x} \! \int \!{\rm d}^3 \boldsymbol{y} \, (\boldsymbol{x} + \boldsymbol{y})^2 \, e^{-i\boldsymbol{p} \cdot (\boldsymbol{x} + \boldsymbol{y})} C^{AB}_{0}(\tau,\tau/2;\boldsymbol{x},\boldsymbol{y}) \, .
\end{align}
We first focus on the radial derivative.
Using the ground-state-saturated result given in eq.~\eqref{eq:C_tmr_saturation}, one finds
\begin{equation}
\boldsymbol{p}\cdot\nabla_{\boldsymbol{p}} C^{(0),AB}_{0}(\tau,\tau/2;-\boldsymbol{p},\boldsymbol{p}) = \frac{Z_\pi^A Z_\pi^B}{2 E_\pi(\boldsymbol{p})} e^{-E_\pi(\boldsymbol{p})\tau} \left [ \boldsymbol{p}\cdot\nabla_{\boldsymbol{p}} F_\pi(Q^2) + \Gamma_p \right ] \, ,
\end{equation}
where we substituted $p_0^{E} = i E_\pi(\boldsymbol{p})$.

The term \(\Gamma_p\) arises from the derivative acting on the energy-dependent prefactor and is given by
\begin{equation}
\Gamma_p = -\boldsymbol{p}^2 \left [ \frac{1 + \tau E_\pi(\boldsymbol{p})}{E_\pi(\boldsymbol{p})^2} \right ] F_\pi(Q^2) \, .
\end{equation}
Note that no derivatives act on the overlap factors, since these are independent of \(\boldsymbol{p}\) at fixed \(p^2 = - m_\pi^2\).

From \(Q^2 = 4 \boldsymbol{p}^2\), one obtains
\begin{equation}
\boldsymbol{p}\cdot\nabla_{\boldsymbol{p}} F_\pi(Q^2) = 8\boldsymbol{p}^2 \frac{\partial F_\pi(Q^2)}{\partial Q^2}\,.
\end{equation}
Thus, at nonzero \(\boldsymbol{p}^2\), the radial derivative isolates the slope. Substituting the radial position-space moment of the three-point correlation function into this relation gives the direct estimator
\begin{multline} \frac{\partial F_\pi(Q^2)}{\partial Q^2} = \frac{1}{8\boldsymbol{p}^2} \lim_{\tau \to \infty} \bigg [ \frac{2 E_\pi(\boldsymbol{p})}{Z_\pi^A Z_\pi^B} e^{E_\pi(\boldsymbol{p}) \tau} \\ \times \left( -i \int \!{\rm d}^3 \boldsymbol{x} \!\int \!{\rm d}^3 \boldsymbol{y} \, \boldsymbol{p}\cdot(\boldsymbol{x} + \boldsymbol{y})\, e^{-i\boldsymbol{p} \cdot (\boldsymbol{x} + \boldsymbol{y})} C^{AB}_{0}(\tau,\tau/2;\boldsymbol{x},\boldsymbol{y}) \right) - \Gamma_p \bigg ] \,.
\end{multline}

At $\boldsymbol p = \boldsymbol 0$, the radial derivative vanishes and one must instead use the Laplacian. Evaluating the ground-state-saturated relation directly at this point, and using $F_\pi(0)=1$, gives
\begin{equation}
-\nabla^2_{\boldsymbol{p}} C^{(0),AB}_{0}(\tau,\tau/2;-\boldsymbol{p},\boldsymbol{p}) \bigg |_{\boldsymbol{p}=\boldsymbol{0}} = \frac{Z_\pi^A Z_\pi^B}{2m_\pi} e^{-m_\pi\tau} \left [ -\nabla^2_{\boldsymbol{p}} F_\pi(Q^2) \bigg |_{\boldsymbol{p}=\boldsymbol{0}} + \frac{3(1+m_\pi\tau)}{m_\pi^2} \right ] \, .
\end{equation}
The form-factor Laplacian at this point satisfies
\begin{equation}
\nabla^2_{\boldsymbol{p}} F_\pi(Q^2) \bigg |_{\boldsymbol{p}=\boldsymbol{0}} = 24 \frac{\partial F_\pi(Q^2)}{\partial Q^2} \bigg |_{Q^2=0}\,.
\end{equation}
Combining this identity with the Laplacian position-space moment yields the charge radius of the pion along the lines of ref.~\cite{Feng:2019geu}:
\begin{equation}
\langle r_\pi^2 \rangle = \lim_{\tau \to \infty} \left [ \frac{2 m_\pi e^{m_\pi \tau}}{Z_\pi^A Z_\pi^B} \! \int \!{\rm d}^3 \boldsymbol{x} \!\int \!{\rm d}^3 \boldsymbol{y} \, \left ( \frac{\boldsymbol{x} + \boldsymbol{y}}{2} \right )^2 \! C^{AB}_{0}(\tau,\tau/2;\boldsymbol{x},\boldsymbol{y}) - \frac{3}{4m_\pi^2} - \frac{3\tau}{4 m_\pi} \right ] \,.
\end{equation}

An alternative way to remove the terms that are constant or linear in $\tau$ is to combine the back-to-back correlator with the equal-momentum correlator in eq.~\eqref{eq:C4_equal_momenta}. To make this explicit, define
\begin{align}
C_{0,-}^{AB}(\tau,\boldsymbol{p}) &\equiv C_0^{AB}(\tau,\tau/2;-\boldsymbol{p},\boldsymbol{p})\,, & C_{0,+}^{AB}(\tau,\boldsymbol{p}) &\equiv C_0^{AB}(\tau,\tau/2;\boldsymbol{p},\boldsymbol{p})\,.
\end{align}
At \(Q^2=0\), the two correlators have the same energy-dependent prefactor. Their ground-state contributions satisfy
\begin{align}
C_{0,-}^{(0),AB}(\tau,\boldsymbol{p}) &= A(\tau,\boldsymbol{p}) F_\pi(Q^2)\,, & C_{0,+}^{(0),AB}(\tau,\boldsymbol{p}) &= A(\tau,\boldsymbol{p})\,,
\end{align}
where \(A(\tau,\boldsymbol{p})= Z_\pi^A Z_\pi^B e^{-E_\pi(\boldsymbol{p})\tau}/[2E_\pi(\boldsymbol{p})]\).

Taking \(\boldsymbol p\to\boldsymbol 0\) and subtracting the Laplacian of the equal-momentum correlator cancels the derivatives of this shared prefactor and gives
\begin{equation}
-\left[ \nabla_{\boldsymbol p}^2C_{0,-}^{(0),AB}(\tau,\boldsymbol p) -\nabla_{\boldsymbol p}^2C_{0,+}^{(0),AB}(\tau,\boldsymbol p) \right] \bigg|_{\boldsymbol p=\boldsymbol 0} = -24\,A(\tau,\boldsymbol 0) \frac{\partial F_\pi(Q^2)}{\partial Q^2} \bigg|_{Q^2=0}\,.
\end{equation}
Using \((\boldsymbol x+\boldsymbol y)^2-(\boldsymbol x-\boldsymbol y)^2=4\boldsymbol x\cdot\boldsymbol y\) and \(\langle r_\pi^2\rangle=-6\,\partial F_\pi/\partial Q^2|_{Q^2=0}\), the position-space estimator for the charge radius becomes
\begin{equation}
\langle r_\pi^2\rangle = \lim_{\tau\to\infty} \frac{ \int \!{\rm d}^3 \boldsymbol{x} \!\int \!{\rm d}^3 \boldsymbol{y} \, \boldsymbol{x}\cdot\boldsymbol{y}\, C^{AB}_{0}(\tau,\tau/2;\boldsymbol{x},\boldsymbol{y}) }{ \int \!{\rm d}^3 \boldsymbol{x} \!\int \!{\rm d}^3 \boldsymbol{y} \, C^{AB}_{0}(\tau,\tau/2;\boldsymbol{x},\boldsymbol{y}) }\,.
\end{equation}

The discussion in this appendix has, following section~\ref{sec:infinite_volume}, been phrased in terms of infinite-volume position-space correlators. In a finite volume one faces the same general choice described in section~\ref{sec:finite_volume}: either use periodic correlators directly, reconstruct or model the missing long-distance information, or modify the boundary conditions so that the desired momentum dependence is built into the finite-volume correlation functions. This is closely related to the Rome method of ref.~\cite{deDivitiis:2012vs}, though further investigation is needed to completely understand the connection.

\section{Complex momentum-transfer domain} \label{app:q2_domain}

In this appendix we determine the domain of complex $Q^2$ accessible while maintaining convergence of the two independent spatial Laplace transforms. We first consider back-to-back momenta and then general kinematics.

Introduce the complex on-shell four-momenta
\begin{equation}
p=(E_\pi(\boldsymbol p),\boldsymbol p),\qquad p'=(E_\pi(\boldsymbol p'),\boldsymbol p') ,
\end{equation}
where $E_\pi(\boldsymbol p)=\sqrt{m_\pi^2+\boldsymbol p^2}$ for any three-vector $\boldsymbol p$. We use the mostly-positive metric, which leads to the on-shell conditions
\begin{equation}
\begin{aligned} p^2 &= \vec{p}^2 - E_\pi(\vec{p})^2 = -m_\pi^2 \,, \\ p'^2 &= {\vec{p}'}^2 - E_\pi(\vec{p}')^2 = -m_\pi^2\,. \end{aligned}
\end{equation}
The invariant $Q^2$, defined in eq.~\eqref{eq:q2_definition} of the main text, is then given by
\begin{equation}
Q^2=(p'-p)^2.
\end{equation}
The independent Laplace transforms require
\begin{equation}
\lVert\Im\boldsymbol p\rVert<m_\pi,\qquad \lVert\Im\boldsymbol p'\rVert<m_\pi \, . \label{eq:app_convergence_bounds}
\end{equation}
These bounds are sufficient for the pion three-point function with the isovector vector current considered here. More generally, any current channel whose lightest contribution has mass greater than or equal to $2m_\pi$ imposes no stronger condition than the two individual pion bounds. The important exception is a current that couples to a single pion, as occurs for the isovector axial current. If the one-pion contribution is isolated and subtracted, the next current-channel contribution begins at $3m_\pi$, and the original convergence region is recovered. Without such a subtraction, the one-pion falloff reduces the symmetric bounds to $\lVert\Im\boldsymbol p\rVert<m_\pi/2$ and $\lVert\Im\boldsymbol p'\rVert<m_\pi/2$. The isoscalar axial current does not produce this exception: it has no single-pion or two-pion contribution, and its heavier $\eta/\eta'$, $f_1$, and four-pion channels do not tighten the pion bounds. Any vacuum contribution must likewise be subtracted before applying the transform.

\subsection{Back-to-back kinematics}

For $\boldsymbol p=-\boldsymbol p'=\boldsymbol k+i\boldsymbol\kappa$, the two energies are equal and
\begin{equation}
Q^2 = 4 \boldsymbol p^2 = 4(\boldsymbol k^2-\boldsymbol\kappa^2) +8i\,\boldsymbol k\cdot\boldsymbol\kappa \,.
\end{equation}
Writing $X=\Re Q^2$ and $Y=\Im Q^2$, the Cauchy--Schwarz inequality gives
\begin{equation}
Y^2\leq64\boldsymbol k^2\boldsymbol\kappa^2 =16\boldsymbol\kappa^2(X+4\boldsymbol\kappa^2)\,.
\end{equation}
In this case, the independent conditions in eq.~\eqref{eq:app_convergence_bounds} reduce to $\vec{\kappa}^2 < m_\pi^2$, and the right-hand side is maximal as $\boldsymbol\kappa^2\to m_\pi^2$. It follows that the back-to-back domain is necessarily bounded by
\begin{equation}
(\Im Q^2)^2<16m_\pi^2(\Re Q^2+4m_\pi^2)\,. \label{eq:back_to_back_domain}
\end{equation}
The border, defined by equating the two sides, is a parabola in the complex $Q^2$ plane that is symmetric about the real axis, opens along the positive real direction, and has its vertex at $Q^2=-4m_\pi^2$. The accessible domain is the interior of this parabola, as illustrated in figures~\ref{fig:summary}(b), \ref{fig:Q2_domain}(b), and \ref{fig:q2_domain_samples}(a).

Conversely, every point satisfying eq.~\eqref{eq:back_to_back_domain} can be reached with collinear $\boldsymbol k$ and $\boldsymbol\kappa$. For a target $Q^2=X+iY$, define $r=\sqrt{X^2+Y^2}$, choose an arbitrary real unit vector $\hat{\boldsymbol n}$, and set
\begin{equation}
\boldsymbol k=\sqrt{\frac{r+X}{8}}\,\hat{\boldsymbol n}, \qquad \boldsymbol\kappa=\operatorname{sgn}(Y) \sqrt{\frac{r-X}{8}}\,\hat{\boldsymbol n}, \label{eq:back_to_back_construction}
\end{equation}
where $\operatorname{sgn}(0)$ may be chosen to be $1$. This gives
\begin{equation}
4(\boldsymbol k^2-\boldsymbol\kappa^2)=X,\qquad 8\boldsymbol k\cdot\boldsymbol\kappa=Y.
\end{equation}
Moreover, the convergence condition follows directly from
\begin{equation}
\boldsymbol\kappa^2<m_\pi^2 \quad\Longleftrightarrow\quad r<X+8m_\pi^2 \quad\Longleftrightarrow\quad Y^2<16m_\pi^2(X+4m_\pi^2) \,.\end{equation}

Thus eq.~\eqref{eq:back_to_back_construction} provides an explicit momentum pair for every point in the parabolic domain. This construction is not unique. For example, one may choose any $t$ satisfying $(r-X)/8<t<m_\pi^2$, and set
\begin{equation}
\boldsymbol\kappa^2=t,\qquad \boldsymbol k^2=\frac{X}{4}+t,\qquad \cos\phi=\frac{Y}{8\sqrt{t(X/4+t)}} \,,
\end{equation}
where $\phi$ is the angle between $\boldsymbol k$ and $\boldsymbol\kappa$. The resulting non-collinear momenta give the same $Q^2=X+iY$, while eq.~\eqref{eq:back_to_back_construction} is recovered at the lower endpoint.

\subsection{General kinematics}

General incoming and outgoing momenta give a larger domain. The equal-mass on-shell conditions first imply
\begin{equation}
Q^2+4m_\pi^2=-(p+p')^2\,. \label{eq:q2_sum_identity}
\end{equation}
To obtain a necessary bound, we write $p=R+iI$, where $R=(\epsilon,\boldsymbol k)$ and $I=(\eta,\boldsymbol\kappa)$ are real. We use an analogous decomposition for the primed quantities.

The on-shell condition on $p$ gives
\begin{equation}
R^2-I^2 = -m_\pi^2,\qquad R\cdot I=0,\qquad R^2 = -m_\pi^2 + \boldsymbol\kappa^2 - \eta^2 \equiv -\mu^2\,, \label{eq:three-eqs}
\end{equation}
and in the final equation we have used eq.~\eqref{eq:app_convergence_bounds} to note that $ -m_\pi^2 + \boldsymbol\kappa^2 < 0$, and thus $R^2 < 0$ can be identified with $- \mu^2$ where $\mu$ is real and non-negative. In fact $R$ is future timelike, additionally satisfying $\epsilon > 0$. This follows since the analytic continuation for $E_\pi(\boldsymbol p)$ is defined on the principal sheet for which $\Re E_\pi(\boldsymbol p) > 0$. The final equation of \eqref{eq:three-eqs} then also leads to the bound $\mu>\lvert\eta\rvert$. Naturally all of the same statements hold for the primed quantities.

Since $R$ and $R'$ are both future timelike, the Minkowski Cauchy--Schwarz inequality gives $R\cdot R'\leq-\mu\mu'$. Equivalently, their sum obeys the reverse triangle inequality
\begin{align}
(R+R')^2 &=R^2+R'^2+2R\cdot R' \\ &\leq-\mu^2-\mu'^2-2\mu\mu' =-(\mu+\mu')^2\,.
\end{align}
For the imaginary parts, the spatial contribution can only increase the Minkowski norm, and the ordinary triangle inequality gives
\begin{align} (I+I')^2 &=\lVert\boldsymbol\kappa+\boldsymbol\kappa'\rVert^2 - (\eta+\eta')^2 \\ &\geq-(\eta+\eta')^2 \geq-(\lvert\eta\rvert+\lvert\eta'\rvert)^2\,.
\end{align}

Combining these two bounds yields
\begin{align}
-\Re(p+p')^2 &=- (R+R')^2 + (I+I')^2\\ &\geq (\mu+\mu')^2-(\lvert\eta\rvert+\lvert\eta'\rvert)^2>0\,.
\end{align}
The final inequality is strict because $\mu+\mu'>\lvert\eta\rvert+\lvert\eta'\rvert$. Together with eq.~\eqref{eq:q2_sum_identity}, this proves the necessary condition $\Re Q^2>-4m_\pi^2$\,.

It remains to show that every point in this half-plane is attainable. It is enough to use collinear complex rapidities, defining
\begin{equation}
\begin{aligned} p &= m_\pi(\cosh\theta,\sinh\theta\,\hat{\boldsymbol z})\,, \qquad & \theta&=u+iv\,, \\ p' &= m_\pi(\cosh\theta',\sinh\theta'\,\hat{\boldsymbol z})\,, \qquad & \theta'&=u'+iv'\,. \end{aligned}
\end{equation}
These parametrizations automatically enforce the on-shell condition $p^2=p'^2=-m_\pi^2$. Expanding the spatial component of $p$ gives
\begin{equation}
\boldsymbol p =m_\pi\sinh(u+iv)\,\hat{\boldsymbol z} =m_\pi\bigl(\sinh u\cos v+i\cosh u\sin v\bigr)\hat{\boldsymbol z}\,,
\end{equation}
and it follows that
\begin{equation}
\lVert\Im\boldsymbol p\rVert =m_\pi\lvert\cosh u\sin v\rvert \,.
\end{equation}
Thus the original convergence bound $\lVert\Im\boldsymbol p\rVert<m_\pi$ in eq.~\eqref{eq:app_convergence_bounds} becomes
\begin{equation}
\lvert\cosh u\sin v\rvert<1\,.
\end{equation}
Within the rapidity band connected to the real axis, $\lvert v\rvert<\pi/2$, this is equivalent to
\begin{equation}
\lvert v\rvert< \alpha(u) \,, \label{eq:v-constraint}
\end{equation}
where we have defined
\begin{equation}
\alpha(u)\equiv\arcsin(\operatorname{sech}u) \,.
\end{equation}

\begin{figure}[t] \centering \includegraphics{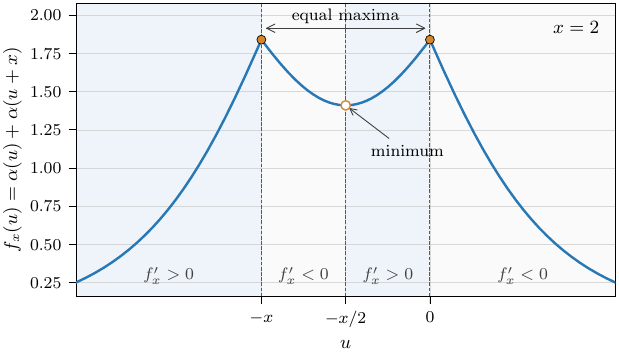} \caption{Illustration of $f_x(u)=\alpha(u)+\alpha(u+x)$ for fixed $x=2$. The dashed vertical lines separate the four monotonicity regions used in the maximization. The filled points at $u=-x$ and $u=0$ mark the equal global maxima, while the open point at $u=-x/2$ marks the local minimum. The signs of $f_x'(u)$ are indicated in each region.} \label{fig:rapidity_bound_maximization} \end{figure}

We next consider the rapidity difference $\theta'-\theta = (u'-u) + i (v'-v) \equiv x+iy$. To work out the full set of attainable $Q^2$ values, it is necessary to first identify the largest possible value of $\lvert y \rvert$ given the constraint \eqref{eq:v-constraint} and its primed analogue. To this end note that, at fixed $u$ and $x$, the individual bounds $\lvert v\rvert<\alpha(u)$ and $\lvert v'\rvert<\alpha(u+x)$ imply
\begin{equation}
\lvert y\rvert = \lvert v'-v \rvert <\alpha(u)+\alpha(u+x)\,.
\end{equation}
The largest allowed value of $\lvert y\rvert$ at fixed $x$ is obtained by maximizing the right-hand side over the common real boost $u$. The function $\alpha$ is even, and, away from the origin,
\begin{equation}
\alpha'(u)=-\operatorname{sgn}(u)\operatorname{sech}u\,.
\end{equation}
It is therefore sufficient to take $x\geq0$. For $u<-x$ the sum $\alpha(u)+\alpha(u+x)$ is increasing, while for $u>0$ it is decreasing. In the interval $-x<u<0$, its derivative is $\operatorname{sech}u-\operatorname{sech}(u+x)$, which is negative for $-x<u<-x/2$ and positive for $-x/2<u<0$. The sum consequently has equal global maxima at $u=0$ and $u=-x$, with a local minimum at $u=-x/2$. These four monotonicity regions and the resulting extrema are illustrated for $x=2$ in figure~\ref{fig:rapidity_bound_maximization}. Thus
\begin{equation}
\max_{u\in\mathbb R}\bigl[\alpha(u)+\alpha(u+x)\bigr] =\alpha(0)+\alpha(x) =\frac{\pi}{2}+\arcsin(\operatorname{sech}x)\,.
\end{equation}
For negative $x$, evenness gives the same result with $x$ replaced by $\lvert x\rvert$. Moreover, the bound is sufficient as well as necessary: choosing $u=0$ and $u'=x$ allows every $y$ strictly inside the resulting interval through suitable choices of $v$ and $v'$. Thus the allowed rapidity differences are those satisfying the bound
\begin{equation}
\lvert y\rvert<\frac{\pi}{2}+\arcsin(\operatorname{sech}\lvert x\rvert)\,. \label{eq:rapidity_difference_domain}
\end{equation}

This result finally yields an explicit construction for every $Q^2$ satisfying $\Re Q^2>-4m_\pi^2$. In terms of $x$ and $y$, the momentum transfer is
\begin{equation}
Q^2=2m_\pi^2[\cosh(x+iy)-1]\,,
\end{equation}
and in this representation, the real and imaginary parts are given by
\begin{align}
X\equiv\Re Q^2&=2m_\pi^2(\cosh x\cos y-1)\,,\\ Y\equiv\Im Q^2&=2m_\pi^2\sinh x\sin y\,.
\end{align}
For a target $Q^2=X+iY$, define
\begin{equation}
a=1+\frac{X}{2m_\pi^2},\qquad b=\frac{Y}{2m_\pi^2},\qquad \rho_\pm=\sqrt{(a\pm1)^2+b^2}.
\end{equation}
An explicit solution of the two relations above is
\begin{equation}
x=\operatorname{arcosh}\!\left(\frac{\rho_++\rho_-}{2}\right),\qquad y=\operatorname{sgn}(b)\arccos\!\left(\frac{\rho_+-\rho_-}{2}\right), \label{eq:general_rapidity_construction}
\end{equation}
where $\operatorname{sgn}(0)$ may be chosen to be $1$. With this choice $x\geq0$ and $\lvert y\rvert\leq\pi$, so the bound in eq.~\eqref{eq:rapidity_difference_domain} is equivalent to
\begin{align}
\lvert y\rvert &<\frac{\pi}{2}+\arcsin(\operatorname{sech}x) \nonumber\\ &\Longleftrightarrow\quad \cos y>-\operatorname{sech}x \quad\Longleftrightarrow\quad a>-1 \quad\Longleftrightarrow\quad X>-4m_\pi^2\,.
\end{align}
Thus eq.~\eqref{eq:general_rapidity_construction} gives an allowed rapidity difference for every point in the open half-plane. To construct the individual rapidities explicitly, set $u=0$ and $u'=x$, and choose
\begin{equation}
v'=\frac{1}{2}\left[\max\!\left(-\alpha(x),y-\frac{\pi}{2}\right) +\min\!\left(\alpha(x),y+\frac{\pi}{2}\right)\right], \qquad v=v'-y.
\end{equation}
Equation~\eqref{eq:rapidity_difference_domain} guarantees that the two intervals overlap, so that $\lvert v\rvert<\pi/2$ and $\lvert v'\rvert<\alpha(x)$. The resulting momenta therefore satisfy both convergence bounds. Together with the necessary condition above, this proves that the full domain generated by general kinematics is precisely $\Re Q^2>-4m_\pi^2$.

As a numerical example, consider the target $Q^2/m_\pi^2=100i$. In this case $a=1$, $b=50$, $\rho_+=\sqrt{2504}$, and $\rho_-=50$, so eq.~\eqref{eq:general_rapidity_construction} gives
\begin{equation}
x\simeq4.605470\,,\qquad y\simeq1.550803\,.
\end{equation}
Choosing $u=0$, $u'=x$, $v'=0$, and $v=-y$ yields the explicit four-momenta
\begin{align}
\frac{p}{m_\pi} &=\bigl(\cos y,-i\sin y\,\hat{\boldsymbol z}\bigr) \simeq\bigl(0.019992,-0.999800i\,\hat{\boldsymbol z}\bigr)\,,\\ \frac{p'}{m_\pi} &=\bigl(\cosh x,\sinh x\,\hat{\boldsymbol z}\bigr) \simeq\bigl(50.019992,50.009995\,\hat{\boldsymbol z}\bigr)\,.
\end{align}
Thus $\lVert\Im\boldsymbol p\rVert/m_\pi\simeq0.999800<1$ and $\lVert\Im\boldsymbol p'\rVert/m_\pi=0$, so both transforms converge, while direct substitution gives $(p'-p)^2/m_\pi^2\simeq100i$ as required. Although formally allowed, such kinematics would likely be highly impractical in a realistic calculation. The imaginary component of the incoming spatial momentum lies very close to the convergence bound, leaving the spatial tails only weakly suppressed after the Laplace weighting and thereby making the estimator particularly susceptible to finite-volume effects and signal-to-noise degradation. At the same time, the very large real spatial momentum of the outgoing state would itself be expected to lead to severe signal-to-noise degradation.

\begin{figure}[t] \centering \includegraphics{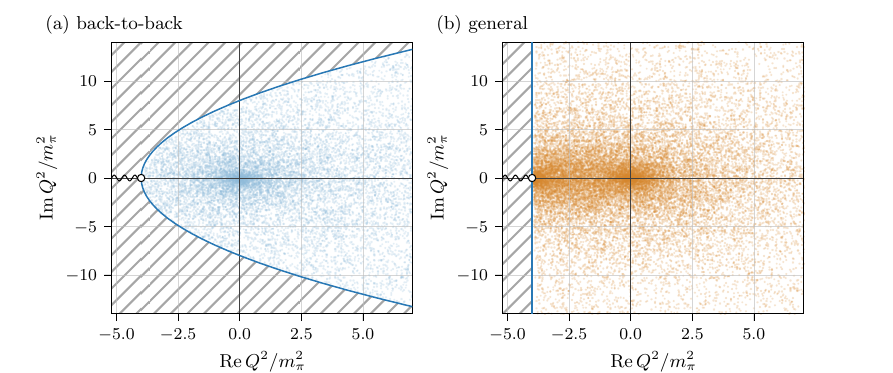} \caption{Numerical illustration of the accessible momentum-transfer domains, with all quantities in units of $m_\pi$. \textbf{(a)} We draw $18\,000$ back-to-back samples with collinear $\boldsymbol k$ and $\boldsymbol\kappa$, taking $k/m_\pi$ uniformly from $[-2.2,2.2]$ and $\kappa/m_\pi$ uniformly from $[-0.995,0.995]$. The samples fill the parabolic domain in eq.~\eqref{eq:back_to_back_domain}. \textbf{(b)} We draw $30\,000$ collinear rapidity pairs with $u=0$, $u'=x$, taking $x$ uniformly from $[-3.2,3.2]$ and $y$ uniformly within $99.5\%$ of the bound in eq.~\eqref{eq:rapidity_difference_domain}. For each pair, $v'$ is chosen at the midpoint of its allowed overlap and $v=v'-y$, ensuring that both individual convergence bounds are satisfied. The resulting samples fill the half-plane $\Re Q^2>-4m_\pi^2$. Solid blue lines show the analytic boundaries, hatching marks inaccessible regions, and the open circle and oscillating line indicate the two-pion branch point and cut.} \label{fig:q2_domain_samples} \end{figure}

The momentum bounds in eq.~\eqref{eq:app_convergence_bounds} are imposed in the Euclidean frame in which the transforms are evaluated, but their image in the invariant $Q^2$ plane has this frame-independent form. Figure~\ref{fig:q2_domain_samples} illustrates both domains with samples that explicitly satisfy the convergence bounds.

 \clearpage

\bibliographystyle{JHEP} \bibliography{ref}

\end{document}